\documentclass[journal=ancac3,manuscript=article,layout=twocolumn]{achemso}

\setkeys{acs}{maxauthors=20, etalmode=truncate}
\usepackage[section]{placeins}
\usepackage{graphicx}
\usepackage{epstopdf}
\usepackage{siunitx}
\usepackage{amsmath}
\usepackage{lineno}
\usepackage{xspace}
\usepackage[version=4]{mhchem}
\usepackage[inline]{enumitem}
\usepackage{booktabs}
\usepackage{threeparttable}
\newcommand{\ie}{\textit{i.e.},\xspace}
\newcommand{\eg}{\textit{e.g.},\xspace}
\newcommand{\etal}{\textit{et al.}}

\newcommand{\mol}[1]{\SI{#1}{M}}

\newcommand{\us}[1]{\SI{#1}{\micro\second}}
\newcommand{\ns}[1]{\SI{#1}{\nano\second}}
\newcommand{\ps}[1]{\SI{#1}{\pico\second}}
\newcommand{\fs}[1]{\SI{#1}{\femto\second}}

\newcommand{\nm}[1]{\SI{#1}{\nano\meter}}

\newcommand{\kPa}[1]{\SI{#1}{\kilo\pascal}}

\newcommand{\K}[1]{\SI{#1}{\kelvin}}
\newcommand{\hour}[1]{\SI{#1}{\hour}}

\newcommand{\nmsq}[1]{\SI{#1}{\nano\meter\squared}}

\graphicspath{{figs/}}
\usepackage{xr}
\makeatletter
\newcommand*{\addFileDependency}[1]{
  \typeout{(#1)}
  \@addtofilelist{#1}
  \IfFileExists{#1}{}{\typeout{No file #1.}}
}
\makeatother

\newcommand*{\myexternaldocument}[1]{
    \externaldocument{#1}
    \addFileDependency{#1.tex}
    \addFileDependency{#1.aux}
}

\myexternaldocument{SI}
\author{Changjiang Liu}
\affiliation[University of Michigan]{Biophysics Program, University of Michigan, Ann Arbor, MI 48109-2125, USA}

\author{Paolo Elvati}
\affiliation[University of Michigan]{Department of Mechanical Engineering, University of Michigan, Ann Arbor, MI 48109-2125, USA}

\author{Angela Violi}
\email{avioli@umich.edu}
\affiliation[University of Michigan]{Biophysics Program, University of Michigan, Ann Arbor, MI 48109-2125, USA}
\alsoaffiliation[University of Michigan]{Department of Mechanical Engineering, University of Michigan, Ann Arbor, MI 48109-2125, USA}
\alsoaffiliation[University of Michigan]{Department of Chemical Engineering, University of Michigan, Ann Arbor, MI 48109-2125, USA}

\title{Antiviral Drug-Membrane Permeability: the Viral Envelope and Cellular Organelles}

\keywords{
lipid bilayer,
kinetic,
permeation,
molecular dynamics,
graphene quantum dots.}

\begin{document}






\makeatletter
\setlength\acs@tocentry@height{9 cm}
\setlength\acs@tocentry@width{4 cm}
\makeatother
\begin{tocentry}
\includegraphics[width=90mm]{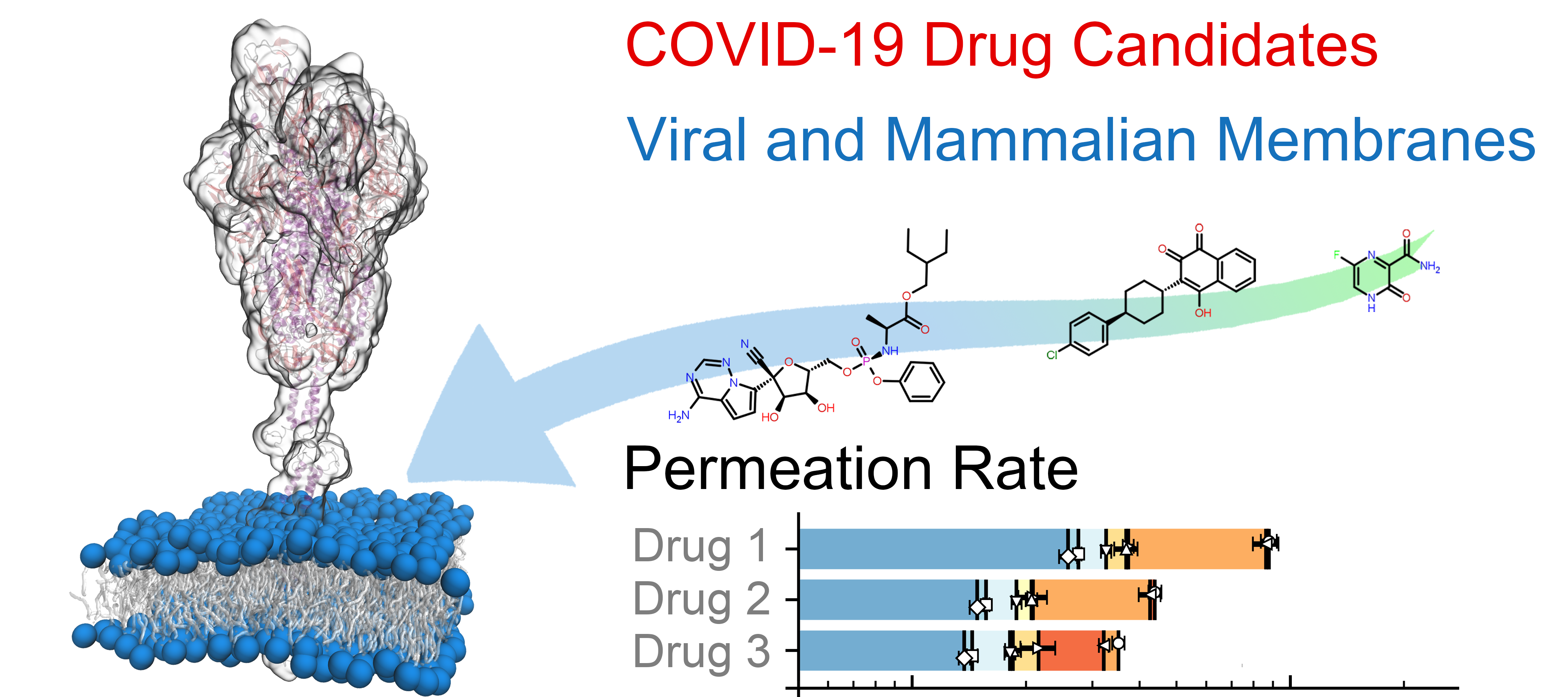}
\end{tocentry}

\begin{abstract}
To shorten the time required to find effective new drugs, like antivirals, a key parameter to consider is membrane permeability, as a compound intended for an intracellular target with poor permeability will have low efficacy. Here, we present a computational model that considers both drug characteristics and membrane properties for the rapid assessment of drugs permeability through the coronavirus envelope and various cellular membranes. We analyze 79 drugs that are considered as potential candidates for the treatment of SARS-CoV-2 and determine their time of permeation in different organelle membranes grouped by viral baits and mammalian processes. The computational results are correlated with experimental data, present in the literature, on bioavailability of the drugs, showing a negative correlation between fast permeation and most promising drugs. 
This model represents an important tool capable of evaluating how  permeability affects the ability of compounds to reach both intended and unintended intracellular targets in an accurate and rapid way. The method is general and flexible and can be employed for a variety of molecules, from small drugs to nanoparticles, as well to a variety of biological membranes.
\end{abstract}

\bigbreak


\section*{Introduction}

\begin{figure*}[!ht]
\centering\includegraphics[width=145mm]{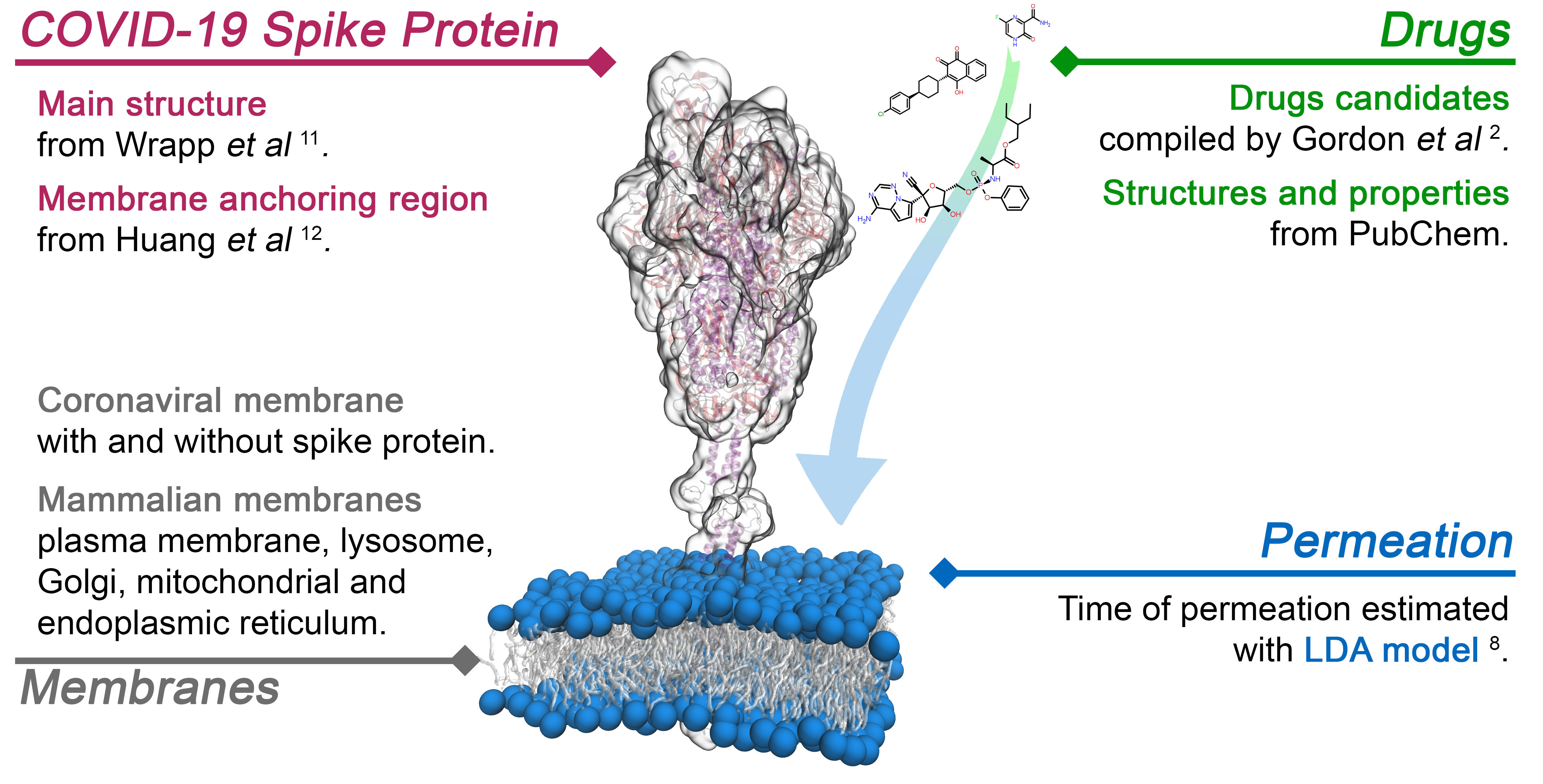}
\caption{Flow and data source used in this work.
}
\label{fig:graph}
\end{figure*}

The 2020 pandemic has brought the chasm between the rapid spread of infectious diseases and the relative slow pace of scientific discovery to stark reality.  This particular event is not the first, nor it will be the last. Since the turn of the century three different coronavirus threats have surfaced and emerging infectious diseases will continue to be a threat for the foreseeable future. 
Current methods for studying betacoronaviruses, responsible for the 2020 pandemic, are technically and time demanding. Viral isolation from field sample is rarely successful and reverse genetics recovery of recombinant virus is labor intensive and expensive. Similarly, the design and testing of new drugs is an extremely lengthy process. To this end, drug repurposing, where existing compounds that have already been tested safe in humans, are redeployed to combat another disease, has emerged as a critical strategy. 
As an example, Remdesivir, one of the promising COVID-19 drug candidates at the moment, is a broad-spectrum drug that was originally targeted toward a number of viruses, such as SARS-CoV and MERS-CoV, but also studied to counter Ebola virus infection.

Multiple characteristics define the potency of a drug, as well as its side effects.
Among them, bioavailability - the fraction of a compound that reaches the intended target - is one of the hardest to estimate reliably and rapidly.
Although several factors contribute to in defining bioavailability, passive permeability through different biological membranes is a critical component since independent of the ability of a drug to bind to its target, the inability to reach the intended target often translates into poor \textit{in vivo} efficacy and may require higher dosages with higher risks of side effects~\cite{aungst1993novel}.

\begin{figure*}[!ht]
\centering\includegraphics[width=135mm]{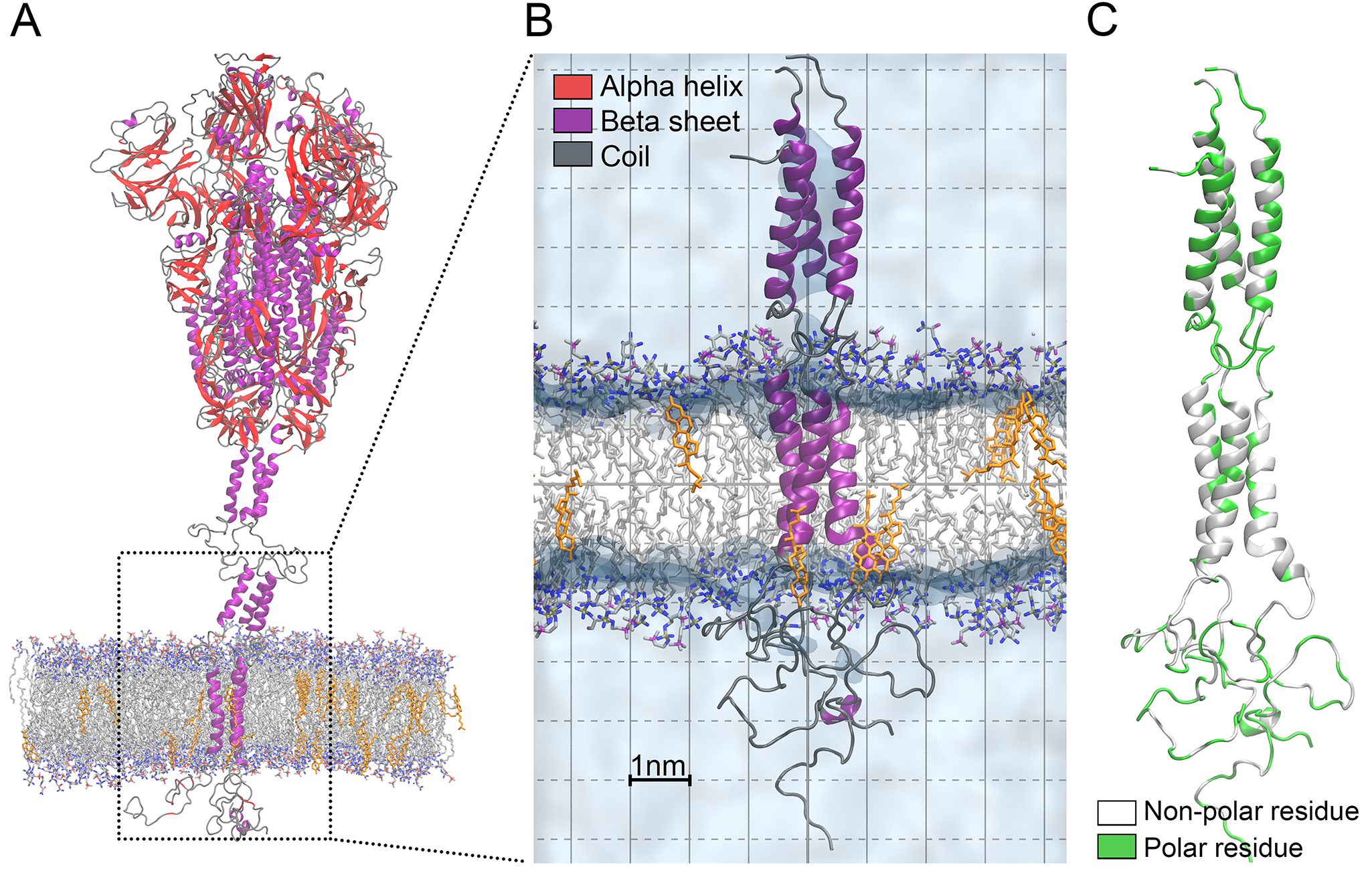}
\caption{
Structural diagram of the coronavirus membrane with the anchored COVID-19 spike protein. 
A) Protein colored according to secondary structure (purple for alpha helix, red for beta sheet, and gray for random coils).
In the lipid bilayer, carbon are shown in gray, nitrogen in red, oxygen in blue, and  phosphate in tan; cholesterol is shown uniformly in orange. Hydrogen atoms omitted for clarity.
B) The anchoring region (part of the heptad repeated, the transmembrane region, and the C-tails) of the spike protein in the membrane.
C) The anchoring region of the spike protein colored by the residue type.
Non-polar residues are colored in white, while polar residues are in green.
}
\label{fig:struct}
\end{figure*}

The viral life cycle encompasses several crucial steps, starting with the attachment of the virus to the cell and finishing with the release of the progeny virions from the host cell. 
SARS-CoV-2 targets cells through the viral structural spike protein that binds to an enzyme on the host cell membranes. Following receptor binding, the virus uses host cell receptors and endosomes to enter cells. Once inside, viral polyproteins are synthesized that encode for the replicase-transcriptase complex.
The virus then synthesizes RNA and structural proteins leading to completion of assembly and release of viral particles. 
These viral lifecycle steps provide potential targets for drug therapy and most of the drugs currently in clinical trials inhibit key components of the infection life-cycle, such as viral entry into the host cell, viral replication, and viral RNA synthesis~\cite{Gordon2020Nature}.
These targets are localized to particular subcellular compartments, yet current drug design strategies are focused on general bioavailability and rarely address drug delivery to specific tissues or intracellular compartments.
Therefore, knowledge of the permeability of antiviral drugs, with reference to both viral and mammalian cell compartments, is vitally important from both a pharmacokinetics and drug design standpoint.

In an effort to accelerate translation of promising drug candidates, as well as eliminate potentially ineffective or even harmful candidate, we present a science-based model to predict the permeation of drug candidates for SARS-CoV-2 into coronavirus membrane and mammalian membranes.

Over the past decades many models have been developed to estimate drugs' permeability.
The most commonly used ones are the empirical "rule of five"~\cite{lipinski1997experimental}, QSAR (Quantitative Structure-Activity Relationship)~\cite{kirchmair2011development} and QSPR (Quantitative Structure-Property Relationship) methods~\cite{2015_Wu_QuantitativeStructurePropertyRelationship}.
While QSAR/QSPR methods have shown their efficacy in screening the binding of anti-viral drugs and targets, the predicted permeability is at best only accurate to an order of magnitude, mostly due to the incorporation of too many descriptors that are not related to the permeation problem, and the strong dependence on the quality of the training sets employed~\cite{gozalbes2011qsar}.
On the other end of the accuracy spectrum, several methods have been proposed over the years that leverage a physical description of the passive permeation process, like the inhomogeneous solubility diffusion model and state-transition model~\cite{venable2019molecular}.
These models, that currently often involve molecular dynamics (MD) simulations, are either computationally too expensive for a rapid application or suffer from limited accuracy when drugs of various sizes and shapes or membranes with different lipid profiles are involved.

To bridge the gap between accuracy and ease of use, we have recently introduced a model to predict the rates of passive permeation of molecules and nanoparticles (up to several \nm{}) into phosphatidylcholine membranes~\cite{Liu2019}.
The Low-Density Area (LDA) model, which can combine both experimental and computational data, has shown accuracy equivalent to fully atomistic MD methods with a much smaller computational cost.  
Such a result was achieved observing that the permeation process of small compounds is controlled by the formation of low-density areas on the membrane surface, and by factorizing the permeation process into independent contributions of the membrane thermal and pressure fluctuations and characteristics of the molecules (size, shape and lipophilicity).

Leveraging the LDA model, here we report on the permeation kinetics of promising drugs candidates through coronavirus membranes, with the overarching goal of providing inputs to help the design and selection of new or repurposed drugs.
Additionally, since the target of many drugs are located inside mammalian cells, we also report on the permeation kinetics of the same drugs through five mammalian membranes.
A schematic of the overall approach and data sources used in this study is shown in Fig.~1. 
The computations were obtained using a combination of molecular dynamics simulations in conjunction with an augmented LDA model for membranes. 
Seventy-nine drugs for SARS-CoV-2 were analyzed, that comprise FDA approved drugs, compounds in clinical trials, and preclinical compounds~\cite{Gordon2020Nature}.
The computational results are compared with data available in the literature on bioavailability and show the potential of the model to aid in the assessment of drugs' permeability into a variety of organelles. 

\section*{Results and Discussion}
\subsection*{Molecular description of coronavirus envelope}

Coronaviridae is a family of large, enveloped, positive-stranded RNA viruses. 
The envelope is derived from the host Endoplasmic Reticulum (ER) membrane and consists of a lipid bilayer, in which structural proteins are anchored. 
The surface glycoprotein spike that forms large protrusions from the virus surface, binds to the host-cell receptor and mediates viral entry~\cite{Belouzard2012,Li2016}. 

Since surface characteristics of the biological membranes determine the permeability of drugs, we developed a model to describe the anchoring of the spike protein to the lipid bilayer. 
The spike contains three segments: a large ectodomain, a single-pass transmembrane anchor, and a short intracellular tail.
The region outside the viral core forms alpha helices where coiled-coil interactions, through hydrophobic contact, help stabilize the chains. 
The tails near the C-terminal are mostly polar with charged residues that form random coils inside the viral core (\ie the cytoplasmic side after membrane fusion with the host cell): interactions between the charged residues and lipid head groups stabilize the end of the transmembrane region of the spike protein.
A diagram of a section of the coronavirus envelope composed of lipid bilayer and the anchored spike protein is shown in Fig.~\ref{fig:struct}.  

Based on existing data and estimated secondary structure~\cite{Wrapp2020,Huang2020}, we determined that the transmembrane region of the COVID-19 spike protein is formed by the residues in the \#1210-1235 range.  
The charge-charge interactions among residues 1215:TYR, 1219:GLY, and 1223:GLY contribute to hold the three chains together inside the hydrophobic region of the membrane lipids.
The density of the spike protein, used in our simulations, reflects the experimental density of about \nmsq{100} to \nmsq{200} per  transmembrane proteins (of any kind), estimated from a viral envelop diameter of \nm{80} to \nm{120}~\cite{Haan2005} and an approximate number of 800 transmembrane proteins per virion~\cite{Hogue1986}.
To reduce the computational cost, after the initial equilibration run used to validate the stability of the protein anchoring region, we considered only part of the protein (Fig.~\ref{fig:struct}B and C) as the rest of the spike protein is far enough (\nm{5} or more) to affect the formation and life time of low-density areas on the viral membrane surface. 

For the lipid bilayer, slightly higher concentrations of sphingomyelin and phosphatidylinositol were found in the viral envelope, compared to the ER membrane from where coronaviruses acquire their membrane envelope~\cite{VanGenderen1995}.

\subsection*{Drugs' Membrane Permeability}

\begin{figure}[!ht]
\centering\includegraphics[width=82mm]{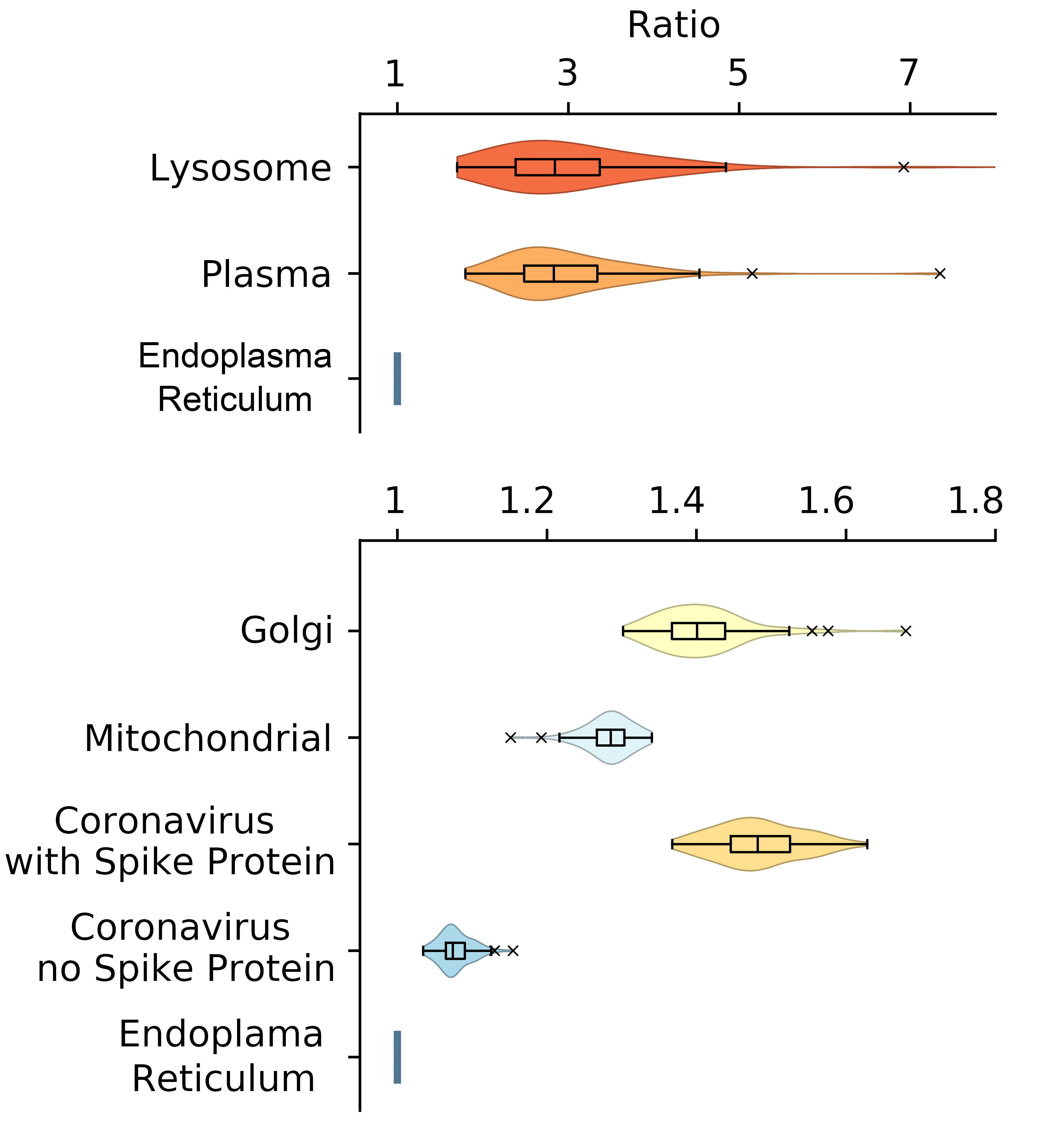}
\caption{
Ratios of times of entry for drugs in different membranes.
The time of entry in ER membrane is chosen as reference.}
\label{fig:compare_membs}
\end{figure}

\begin{figure*}[!ht]
\centering\includegraphics[width=165mm]{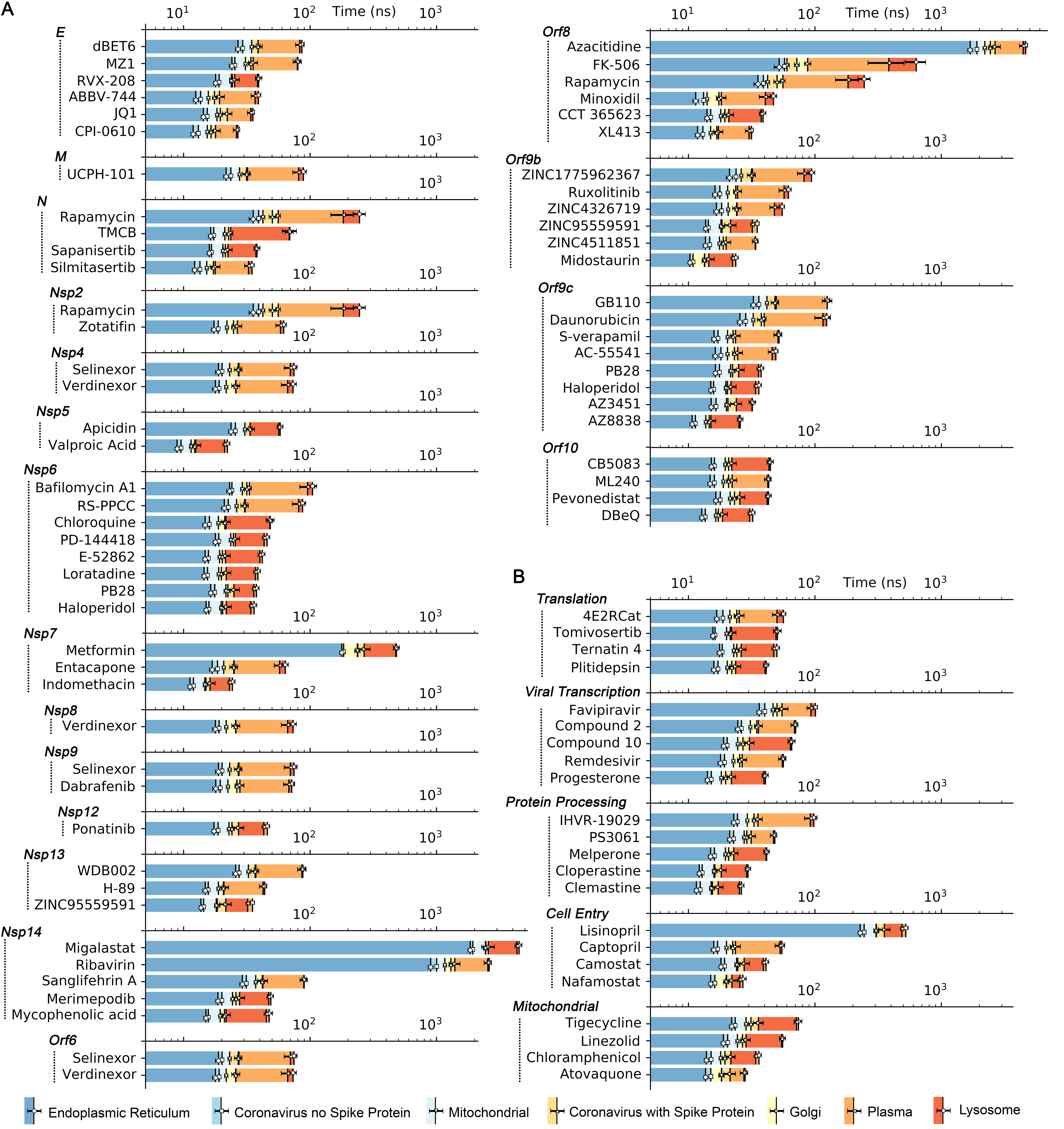}
\caption{
Times of permeation for drugs candidates in different organelle membranes, grouped by A) viral baits, B) related mammalian processes when viral baits were unknown.
Error bars represent one standard deviation.
Tabulated data in the supplementary materials.
}
\label{fig:drugs_data}
\end{figure*}

The LDA model describes the permeation as a gated entry process, regulated by the chance of the molecule to find a low-density area on the membrane surface and by the solubility of the drug in the hydrophobic phase. 
The four parameters used in the model separate the contributions of the characteristics of the molecules (size, shape, solubility) from the properties of the membranes (lipid density distribution).

In addition to the viral envelope described before (with and without spike protein), we studied the surface properties of five membranes corresponding to various mammalian compartments (lysosome, plasma, Golgi, mitochondrial and endoplasmic reticulum membranes) using all-atom molecular dynamics simulations. 

Fig.~\ref{fig:compare_membs} reports the distribution of the permeation times of the 79 drugs in all the membranes relative to the ER membrane, which was chosen as reference since the permeation in this membrane is the fastest for all drugs. 
The results show that the inclusion of the transmembrane protein in the COVID-19 membrane has a large impact on drug's membrane permeability.
The presence of the transmembrane protein affects the frequency, shape, and lifetime of low-density areas in the LDA model. 
Indeed, when comparing the distribution of permeation times for ER and coronavirus membrane, with and without the spike protein, we observe a $\sim$40\% increase in the permeation time due to the presence of proteins and $\sim$10\% due to the change in lipid composition.
This effect is mostly due to the reduced mobility of the lipids near a rigid object similar to what has been observed for C$_{60}$ fullerenes in lipid bilayers~\cite{2016_Russ_C60fullerenelocalization}. 

While the presence of the spike protein in the COVID-19 membrane results in an increased stiffness, the viral membrane is still generally two times more permeable than plasma membrane and lysosomes (before accessible surface and concentration effects are considered).
Although currently very few drugs are required to permeate through viral membranes, this difference can potentially be exploited by other classes of compounds, like nanoparticles, which can amplify the effect of membrane stiffness due to their relative large size.
The individual times of permeation of drug candidates grouped by viral baits or, when that is unknown, by biological processes, are shown in Fig.~\ref{fig:drugs_data}.
The results show a variety of timescales going from nanoseconds to microseconds depending on the drug and membrane. 

Albeit the bioavailability of a drug is affected by many factors, like metabolism rate and plasma protein binding~\cite{khan2016various}, we find a negative correlation between the bioavailability of drugs and the permeation time predicted by the LDA model.
For example, for drugs within the same group, Lisinopril (Fig.~\ref{fig:drugs_data}B, Cell Entry) is the least permeable compound as well as the compound with the lowest bioavailability (25\% based on experimental data~\cite{beermann1988pharmacokinetics}). 
On the other hand, Captopril, Nafamostat, and Camostat, for which we predict permeation times that are one order of magnitude shorter than Lisinopril, have an estimated bioavailability of 100\%~\footnote{data from http://drugbank.ca}.
Similar observations can be made for Rapamycin (Fig.~\ref{fig:drugs_data}A, N), which has lower bioavailability than Sapanisertib and Silmitasertib.
While undeniably other factors affect the ability of a drug to reach the intended target, these results show how useful membrane permeability predictions can be. 

Interestingly, the permeation times of drugs that are currently identified as most promising candidates against COVID-19 during their clinical trial~\cite{Krogan2020} are among the shortest compared to other drugs with similar targets.
For example, drugs related to the disruption of viral translation such as Ternatin-4~\cite{Reich2018} (Fig.~\ref{fig:drugs_data}B Translation), Zotatifin~\cite{thompson2017preclinical} (Fig.~\ref{fig:drugs_data}A, Nsp2) and Plitidepsin~\cite{leisch2019plitidepsin} (Fig.~\ref{fig:drugs_data}B, Translation), and sigma receptors such as Haloperidol~\cite{Hellewell1994} and Compound~PB28~\cite{Azzariti2006} (Fig.~\ref{fig:drugs_data}A Nsp6 and Orf9c), Melperone, Cloperastine~\cite{Cuzzocrea2011}, and Clemastine (Fig.~\ref{fig:drugs_data}B, Protein Processing), and the female hormone progesterone (Fig.~\ref{fig:drugs_data}B, Viral transcription) are among the most permeable compounds in their respective categories.
While a direct comparison among these drugs is not straightforward, as they need to cross different membranes in order to reach their intended target, the time of permeation in their least permeable membrane (\ie lysosome or plasma) is \ns{40} on average with an upper bound of \ns{62}, while the remaining drugs require an average of \ns{242} and a maximum of \us{4.6}.

\section*{Conclusion}

In this paper,  we present a model to compute the permeability of drugs that target various steps in the lifecycle of infection of SARS-CoV-2. The model is based on the LDA theory developed by our group that describes the permeation process as regulated by low-density area on the membrane surface and solubility of drugs in the hydrophobic phase of the membrane itself. 
Starting from the virus structure, we developed an atomistic model of the spike protein anchored to the lipid bilayer to study the viral envelope.  Molecular dynamics simulations of the viral and five mammalian membranes were carried out to determine the dynamics and characteristics of the membranes.  The computational results were compared with data present in the literature on bioavailability, and show that the LDA model can be applied to predict the permeability of drugs through various membranes and to gain insights on the differences between membranes and potentially organs or organisms.
From a drug design, as well as health assessment perspective, this predictive capability can facilitate compound evaluation by ruling out candidates with low permeability or excessive permeability in unintended organs or organelles.
Indeed, the LDA model can be used to estimate the absorption selectivity of different targets (\eg carcinogenic cells, organs) by molecules and nanoparticles without need of large set of preexisting data, making it an invaluable tool for the comprehensive and fast validation pipeline of new compounds. 

While the results reported in this paper focus on drugs for SARS-CoV-2, the method is general and flexible and can be employed, with other tools to shorten the effective response time to a global pandemic. Our work provides a means for scientific discovery to move closer to the speed of pandemic spreads and it can be generalized to future events.

\section*{Methods}

\subsection*{Membrane simulations}
We used NAMD~\cite{Phillips2005} version~2.13 with CHARMM Force Field~\cite{Klauda2010} version~36 for lipids, and TIP3P~\cite{MacKerell1998} for water molecules.
All MD simulations were performed with a timestep of \fs{2} while keeping all the \ce{C-H} and \ce{O-H} bonds rigid \textit{via} the SHAKE algorithm~\cite{Ryckaert1977}.
Long range electrostatic interactions were modeled with the particle mesh Ewald method~\cite{Darden1993} using a \nm{0.1} grid spacing, a tolerance of 10$^{-6}$ and cubic interpolation; cubic periodic boundary conditions were applied.
Temperature was kept constant at \K{310} using a Langevin thermostat~\cite{Loncharich1992} with a time constant of \ps{1}, while pressure was kept at \kPa{101.325} using Langevin piston method~\cite{MartynaTobiasKlein1994,Feller1995} with period of \fs{50} and \fs{25} decay.
To account for the intrinsic anisotropy of the system, the production simulations were performed in the NPsT ensemble, where the x and y dimensions of the periodic system (coplanar with the membrane) are allowed to vary independently from the z dimension.
Non-bonded short-range interactions where smoothly switched to 0 between 1 and \nm{1.2} with a X-PLOR switch function.

The lipid compositions for cellular organelles (plasma, lysosome, Golgi, mitochondrial) were taken from Horvath~\etal~\cite{Horvath2013}, and the ER and viral membrane from van~Genderen~\etal~\cite{VanGenderen1995}.
Table~S1 provides more details about the composition of each membrane.
Membranes without spike proteins had an approximate size of \nm{8} by \nm{8}, subject to round up to satisfy the required imposed by the lipid composition.
The membranes with spike proteins were extended to guarantee at least \nm{4} of separation between the closest atom of the spike protein with its nearest periodic image, resulting in membranes of approximately \nm{15.4} by \nm{15.4} for the system with the entire protein and \nm{9.5} by \nm{9.5} for the system with the anchoring region only.
System were simulated in explicit solvent to guarantee at least \nm{6} of separation between the membrane or protein and their periodic image from top and bottom. 
The lipids, protein, water molecules and ions were initially placed by the CHARMM-GUI membrane builder~\cite{Jo2007}.

Each membrane was equilibrated for at least \ns{30} in the NPsT ensemble described above.
The membranes were assumed to be at equilibrium once the time average of the area per lipid in a NPsT simulations was varying less than 10\% in \ns{10}.
The surface properties (LDAs) were then obtained by running a additional \ns{100}.

\subsection*{Spike protein in membrane}
The main structure (residues 27 to 1146) of the spike protein can be found in the protein data bank \#6VSB, based on the cryo-EM structure of Wrapp \etal~\cite{Wrapp2020}.
The secondary structure of residue 1146 to 1273 was instead based on the prediction of the C-I-TASSER model~\cite{Huang2020}.
The secondary structure of the alpha helices of the lower chunk was left unmodified, while the random coils were twisted to align the alpha helices perpendicularly to the membrane.
Simulations of \ns{30} show that the anchor is stable with no lipid rearrangement or vertical translation relative to the membrane.

\subsection*{Drug properties}
To better factor the drugs' flexibility in the model, their size/shape was estimated by considering a large sample of conformations at room temperature in water.
For this purpose, we run molecular dynamics simulations using the CHARMM General Force Field~\cite{Vanommeslaeghe2010} to model the atomic forces and collected the molecule's conformations every \ps{20}.
CGenFF~\cite{Vanommeslaeghe2012} software was used to assist in the production of the needed topology files.
Each drug molecule was presented at the center of an cubic box of \nm{6} in each dimension, containing explicit solvent of water and \mol{0.15}~\ce{NaCl}.

MD setup for drug simulation was the same as the membrane simulations, except the use of an isotropic ensemble (NPT).
Each system was minimized for 1000 steps, following \ns{0.5} of relaxation and \ns{0.5} of production.
Every \ps{20} the molecule shape was estimated using the protocol described previously~\cite{Liu2019} and then used to calculate the $P_{LDA}$ for the specific conformation.
The final value for $P_{LDA}$ was computed by averaging over the values obtained for all the conformations.
Finally, the partition coefficients of the drugs molecules in octanol and water were taken from PubChem website using experimental values when available, or prediction from XLogP3 model~\cite{Cheng2007} otherwise.

\begin{acknowledgement}

The authors are grateful to Dr. Matteo Masetti and Dr. J. Scott VanEpps for the discussions and suggestions.
This work was supported by the University of Michigan, BlueSky project.

\end{acknowledgement}

\begin{suppinfo}
Additional information about the drugs and membranes involved in this study can be found in the supplementary information.
This material is available for download \textit{via} the Internet at \texttt{http://pubs.acs.org/}.
\end{suppinfo}

\bibliography{main}

\providecommand{\latin}[1]{#1}
\makeatletter
\providecommand{\doi}
  {\begingroup\let\do\@makeother\dospecials
  \catcode`\{=1 \catcode`\}=2 \doi@aux}
\providecommand{\doi@aux}[1]{\endgroup\texttt{#1}}
\makeatother
\providecommand*\mcitethebibliography{\thebibliography}
\csname @ifundefined\endcsname{endmcitethebibliography}
  {\let\endmcitethebibliography\endthebibliography}{}
\begin{mcitethebibliography}{39}
\providecommand*\natexlab[1]{#1}
\providecommand*\mciteSetBstSublistMode[1]{}
\providecommand*\mciteSetBstMaxWidthForm[2]{}
\providecommand*\mciteBstWouldAddEndPuncttrue
  {\def\EndOfBibitem{\unskip.}}
\providecommand*\mciteBstWouldAddEndPunctfalse
  {\let\EndOfBibitem\relax}
\providecommand*\mciteSetBstMidEndSepPunct[3]{}
\providecommand*\mciteSetBstSublistLabelBeginEnd[3]{}
\providecommand*\EndOfBibitem{}
\mciteSetBstSublistMode{f}
\mciteSetBstMaxWidthForm{subitem}{(\alph{mcitesubitemcount})}
\mciteSetBstSublistLabelBeginEnd
  {\mcitemaxwidthsubitemform\space}
  {\relax}
  {\relax}

\bibitem[Aungst(1993)]{aungst1993novel}
Aungst,~B.~J. {Novel Formulation Strategies For Improving Oral Bioavailability
  of Drugs with Poor Membrane Permeation or Presystemic Metabolism.} \emph{J.
  Pharm. Sci} \textbf{1993}, \emph{82}, 979--987\relax
\mciteBstWouldAddEndPuncttrue
\mciteSetBstMidEndSepPunct{\mcitedefaultmidpunct}
{\mcitedefaultendpunct}{\mcitedefaultseppunct}\relax
\EndOfBibitem
\bibitem[Gordon \latin{et~al.}()Gordon, Jang, Bouhaddou, Xu, Obernier, Kris,
  Meara, Rezelj, Guo, Swaney, Tia, Huettenhain, Kaake, Richards, Tutuncuoglu,
  Batra, Haas, Modak, Kim, Haas, Benjamin, Braberg, Fabius, Eckhardt,
  Soucheray, Bennett, Cakir, Mcgregor, Li, Meyer, Vallet, Kain, Miorin, Moreno,
  Zar, Naing, Peng, Shi, Zhang, Shen, Kirby, Melnyk, Neal, Cai, Chang,
  Broadhurst, Klippsten, Sharp, Cavero, Hiatt, Roth, Rathore, Subramanian,
  Noack, Malik, Fujimori, Ideker, Craik, Floor, James, Gross, Sali, Roth,
  Ruggero, Taunton, Beltrao, Vignuzzi, Garc{\'{i}}a-sastre, Shokat, and
  Brian]{Gordon2020Nature}
Gordon,~D.~E.; Jang,~G.~M.; Bouhaddou,~M.; Xu,~J.; Obernier,~K.; Kris,~M.;
  Meara,~M. J.~O.; Rezelj,~V.~V.; Guo,~J.~Z.; Swaney,~D.~L.; Tia,~A.;
  Huettenhain,~R.; Kaake,~R.~M.; Richards,~A.~L.; Tutuncuoglu,~B.; Batra,~J.;
  Haas,~K.; Modak,~M.; Kim,~M.; Haas,~P. \latin{et~al.}  {A SARS-CoV-2 Protein
  Interaction Map Reveals Targets for Drug Repurposing}. \emph{Nature} (Just
  Accepted)\relax
\mciteBstWouldAddEndPuncttrue
\mciteSetBstMidEndSepPunct{\mcitedefaultmidpunct}
{\mcitedefaultendpunct}{\mcitedefaultseppunct}\relax
\EndOfBibitem
\bibitem[Lipinski \latin{et~al.}(1997)Lipinski, Lombardo, Dominy, and
  Feeney]{lipinski1997experimental}
Lipinski,~C.~A.; Lombardo,~F.; Dominy,~B.~W.; Feeney,~P.~J. Experimental and
  Computational Approaches to Estimate Solubility and Permeability in Drug
  Discovery and Development Settings. \emph{Adv. Drug Delivery Rev.}
  \textbf{1997}, \emph{23}, 3--25\relax
\mciteBstWouldAddEndPuncttrue
\mciteSetBstMidEndSepPunct{\mcitedefaultmidpunct}
{\mcitedefaultendpunct}{\mcitedefaultseppunct}\relax
\EndOfBibitem
\bibitem[Kirchmair \latin{et~al.}(2011)Kirchmair, Distinto, Roman~Liedl, Markt,
  Maria~Rollinger, Schuster, Maria~Spitzer, and
  Wolber]{kirchmair2011development}
Kirchmair,~J.; Distinto,~S.; Roman~Liedl,~K.; Markt,~P.; Maria~Rollinger,~J.;
  Schuster,~D.; Maria~Spitzer,~G.; Wolber,~G. Development of Anti-Viral Agents
  Using Molecular Modeling and Virtual Screening Techniques. \emph{Infect.
  Disord.: Drug Targets} \textbf{2011}, \emph{11}, 64--93\relax
\mciteBstWouldAddEndPuncttrue
\mciteSetBstMidEndSepPunct{\mcitedefaultmidpunct}
{\mcitedefaultendpunct}{\mcitedefaultseppunct}\relax
\EndOfBibitem
\bibitem[Wu \latin{et~al.}()Wu, Zhang, Lin, Chen, Guo, Qian, and
  Zhang]{2015_Wu_QuantitativeStructurePropertyRelationship}
Wu,~W.; Zhang,~C.; Lin,~W.; Chen,~Q.; Guo,~X.; Qian,~Y.; Zhang,~L. Quantitative
  {{Structure}}-{{Property Relationship}} ({{QSPR}}) {{Modeling}} of
  {{Drug}}-{{Loaded Polymeric Micelles}} via {{Genetic Function
  Approximation}}. \emph{10}, e0119575\relax
\mciteBstWouldAddEndPuncttrue
\mciteSetBstMidEndSepPunct{\mcitedefaultmidpunct}
{\mcitedefaultendpunct}{\mcitedefaultseppunct}\relax
\EndOfBibitem
\bibitem[Gozalbes \latin{et~al.}(2011)Gozalbes, Jacewicz, Annand, Tsaioun, and
  Pineda-Lucena]{gozalbes2011qsar}
Gozalbes,~R.; Jacewicz,~M.; Annand,~R.; Tsaioun,~K.; Pineda-Lucena,~A.
  QSAR-based permeability model for drug-like compounds. \emph{Bioorg. Med.
  Chem.} \textbf{2011}, \emph{19}, 2615--2624\relax
\mciteBstWouldAddEndPuncttrue
\mciteSetBstMidEndSepPunct{\mcitedefaultmidpunct}
{\mcitedefaultendpunct}{\mcitedefaultseppunct}\relax
\EndOfBibitem
\bibitem[Venable \latin{et~al.}(2019)Venable, Kr{\"{a}}mer, and
  Pastor]{venable2019molecular}
Venable,~R.~M.; Kr{\"{a}}mer,~A.; Pastor,~R.~W. Molecular Dynamics Simulations
  of Membrane Permeability. \emph{Chem. Rev.} \textbf{2019}, \emph{119},
  5954--5997\relax
\mciteBstWouldAddEndPuncttrue
\mciteSetBstMidEndSepPunct{\mcitedefaultmidpunct}
{\mcitedefaultendpunct}{\mcitedefaultseppunct}\relax
\EndOfBibitem
\bibitem[Liu \latin{et~al.}(2019)Liu, Elvati, Majumder, Wang, Liu, and
  Violi]{Liu2019}
Liu,~C.; Elvati,~P.; Majumder,~S.; Wang,~Y.; Liu,~A.~P.; Violi,~A. {Predicting
  the Time of Entry of Nanoparticles in Lipid Membranes}. \emph{ACS Nano}
  \textbf{2019}, \emph{13}, 10221--10232\relax
\mciteBstWouldAddEndPuncttrue
\mciteSetBstMidEndSepPunct{\mcitedefaultmidpunct}
{\mcitedefaultendpunct}{\mcitedefaultseppunct}\relax
\EndOfBibitem
\bibitem[Belouzard \latin{et~al.}(2012)Belouzard, Millet, Licitra, and
  Whittaker]{Belouzard2012}
Belouzard,~S.; Millet,~J.~K.; Licitra,~B.~N.; Whittaker,~G.~R. {Mechanisms of
  Coronavirus Cell Entry Mediated by the Viral Spike Protein}. \emph{Viruses}
  \textbf{2012}, \emph{4}, 1011--1033\relax
\mciteBstWouldAddEndPuncttrue
\mciteSetBstMidEndSepPunct{\mcitedefaultmidpunct}
{\mcitedefaultendpunct}{\mcitedefaultseppunct}\relax
\EndOfBibitem
\bibitem[Li(2016)]{Li2016}
Li,~F. {Structure, Function, and Evolution of Coronavirus Spike Proteins}.
  \emph{Annu. Rev. Virol.} \textbf{2016}, \emph{3}, 237--261\relax
\mciteBstWouldAddEndPuncttrue
\mciteSetBstMidEndSepPunct{\mcitedefaultmidpunct}
{\mcitedefaultendpunct}{\mcitedefaultseppunct}\relax
\EndOfBibitem
\bibitem[Wrapp \latin{et~al.}(2020)Wrapp, Wang, Corbett, Goldsmith, Hsieh,
  Abiona, Graham, and McLellan]{Wrapp2020}
Wrapp,~D.; Wang,~N.; Corbett,~K.~S.; Goldsmith,~J.~A.; Hsieh,~C.~L.;
  Abiona,~O.; Graham,~B.~S.; McLellan,~J.~S. {Cryo-Em Structure of the
  2019-Ncov Spike in the Prefusion Conformation}. \emph{Science} \textbf{2020},
  \emph{367}, 1260--1263\relax
\mciteBstWouldAddEndPuncttrue
\mciteSetBstMidEndSepPunct{\mcitedefaultmidpunct}
{\mcitedefaultendpunct}{\mcitedefaultseppunct}\relax
\EndOfBibitem
\bibitem[Huang \latin{et~al.}(2020)Huang, Pearce, and Zhang]{Huang2020}
Huang,~X.; Pearce,~R.; Zhang,~Y. {Computational Design of Peptides to Block
  Binding of the SARS-CoV-2 Spike Protein to Human ACE2}. \emph{bioRxiv}
  \textbf{2020}, \relax
\mciteBstWouldAddEndPunctfalse
\mciteSetBstMidEndSepPunct{\mcitedefaultmidpunct}
{}{\mcitedefaultseppunct}\relax
\EndOfBibitem
\bibitem[Haan and Rottier(2005)Haan, and Rottier]{Haan2005}
Haan,~C. A. M.~D.; Rottier,~P. J.~M. {Molecular Interactions in the Assembly of
  Coronaviruses}. \emph{Adv. Virus Res.} \textbf{2005}, \emph{64},
  165--230\relax
\mciteBstWouldAddEndPuncttrue
\mciteSetBstMidEndSepPunct{\mcitedefaultmidpunct}
{\mcitedefaultendpunct}{\mcitedefaultseppunct}\relax
\EndOfBibitem
\bibitem[Hogue and Brian(1986)Hogue, and Brian]{Hogue1986}
Hogue,~B.~G.; Brian,~D.~A. {Structural Proteins of Human Respiratory
  Coronavirus Oc43}. \emph{Virus Res.} \textbf{1986}, \emph{5}, 131--144\relax
\mciteBstWouldAddEndPuncttrue
\mciteSetBstMidEndSepPunct{\mcitedefaultmidpunct}
{\mcitedefaultendpunct}{\mcitedefaultseppunct}\relax
\EndOfBibitem
\bibitem[van Genderen \latin{et~al.}(1995)van Genderen, Godeke, Rottier, and
  van Meer]{VanGenderen1995}
van Genderen,~I.~L.; Godeke,~G.~J.; Rottier,~P. J.~M.; van Meer,~G. {The
  Phospholipid Composition of Enveloped Viruses Depends on the Intracellular
  Membrane Through Which They Bud}. \emph{Biochem. Soc. Trans.} \textbf{1995},
  \emph{23}, 523--526\relax
\mciteBstWouldAddEndPuncttrue
\mciteSetBstMidEndSepPunct{\mcitedefaultmidpunct}
{\mcitedefaultendpunct}{\mcitedefaultseppunct}\relax
\EndOfBibitem
\bibitem[Russ \latin{et~al.}()Russ, Elvati, Parsonage, Dews, Jarvis, Ray,
  Schneider, Smith, Williamson, Violi, and
  Philbert]{2016_Russ_C60fullerenelocalization}
Russ,~K.~A.; Elvati,~P.; Parsonage,~T.~L.; Dews,~A.; Jarvis,~J.~A.; Ray,~M.;
  Schneider,~B.; Smith,~P. J.~S.; Williamson,~P. T.~F.; Violi,~A.;
  Philbert,~M.~A. C60 Fullerene Localization and Membrane Interactions in
  {{RAW}} 264.7 Immortalized Mouse Macrophages. \emph{8}, 4134--4144,
  Article\relax
\mciteBstWouldAddEndPuncttrue
\mciteSetBstMidEndSepPunct{\mcitedefaultmidpunct}
{\mcitedefaultendpunct}{\mcitedefaultseppunct}\relax
\EndOfBibitem
\bibitem[Khan and Singh(2016)Khan, and Singh]{khan2016various}
Khan,~A.~D.; Singh,~L. {Various Techniques of Bioavailability Enhancement: a
  Review.} \emph{J. Drug Delivery Ther.} \textbf{2016}, \emph{6}, 34--41\relax
\mciteBstWouldAddEndPuncttrue
\mciteSetBstMidEndSepPunct{\mcitedefaultmidpunct}
{\mcitedefaultendpunct}{\mcitedefaultseppunct}\relax
\EndOfBibitem
\bibitem[Beermann(1988)]{beermann1988pharmacokinetics}
Beermann,~B. {Pharmacokinetics of Lisinopril.} \emph{Am. J. Med.}
  \textbf{1988}, \emph{85}, 25--30\relax
\mciteBstWouldAddEndPuncttrue
\mciteSetBstMidEndSepPunct{\mcitedefaultmidpunct}
{\mcitedefaultendpunct}{\mcitedefaultseppunct}\relax
\EndOfBibitem
\bibitem[Krogan(2020)]{Krogan2020}
Krogan,~N. {We Found and Tested 47 Old Drugs That Might Treat the Coronavirus:
  Results Show Promising Leads and a Whole New Way to Fight Covid-19}. 2020;
  https://theconversation.com/we-found-and-tested-47-old-drugs-that-might-treat-the-coronavirus-results-show-promising-leads-and-a-whole-new-way-to-fight-covid-19-136789\relax
\mciteBstWouldAddEndPuncttrue
\mciteSetBstMidEndSepPunct{\mcitedefaultmidpunct}
{\mcitedefaultendpunct}{\mcitedefaultseppunct}\relax
\EndOfBibitem
\bibitem[Reich \latin{et~al.}(2018)Reich, Sprengeler, Chiang, Appleman, Chen,
  Clarine, Eam, Ernst, Han, Goel, Han, Huang, Hung, Jemison, Jessen, Molter,
  Murphy, Neal, Parker, Shaghafi, Sperry, Staunton, Stumpf, Thompson, Tran,
  Webber, Wegerski, Zheng, and Webster]{Reich2018}
Reich,~S.~H.; Sprengeler,~P.~A.; Chiang,~G.~G.; Appleman,~J.~R.; Chen,~J.;
  Clarine,~J.; Eam,~B.; Ernst,~J.~T.; Han,~Q.; Goel,~V.~K.; Han,~E.~Z.;
  Huang,~V.; Hung,~I.~N.; Jemison,~A.; Jessen,~K.~A.; Molter,~J.; Murphy,~D.;
  Neal,~M.; Parker,~G.~S.; Shaghafi,~M. \latin{et~al.}  {Structure-Based Design
  of Pyridone-Aminal eFT508 Targeting Dysregulated Translation by Selective
  Mitogen-Activated Protein Kinase Interacting Kinases 1 and 2 (mnk1/2)
  Inhibition}. \emph{J. Med. Chem.} \textbf{2018}, \emph{61}, 3516--3540\relax
\mciteBstWouldAddEndPuncttrue
\mciteSetBstMidEndSepPunct{\mcitedefaultmidpunct}
{\mcitedefaultendpunct}{\mcitedefaultseppunct}\relax
\EndOfBibitem
\bibitem[Thompson \latin{et~al.}(2017)Thompson, Eam, Young, Fish, Chen,
  Barrera, Howard, Parra, Molter, Staunton, \latin{et~al.}
  others]{thompson2017preclinical}
Thompson,~P.~A.; Eam,~B.; Young,~N.~P.; Fish,~S.; Chen,~J.; Barrera,~M.;
  Howard,~H.; Parra,~A.; Molter,~J.; Staunton,~J., \latin{et~al.}  Preclinical
  Evaluation of eFT226, a Novel, Potent and Selective eIF4A Inhibitor with
  Anti-Tumor Activity in B-Cell Malignancies. \emph{Blood} \textbf{2017},
  \emph{130}, 1530--1530\relax
\mciteBstWouldAddEndPuncttrue
\mciteSetBstMidEndSepPunct{\mcitedefaultmidpunct}
{\mcitedefaultendpunct}{\mcitedefaultseppunct}\relax
\EndOfBibitem
\bibitem[Leisch \latin{et~al.}(2019)Leisch, Egle, and
  Greil]{leisch2019plitidepsin}
Leisch,~M.; Egle,~A.; Greil,~R. Plitidepsin: a Potential New Treatment For
  Relapsed/Refractory Multiple Myeloma. \emph{Future Oncol.} \textbf{2019},
  \emph{15}, 109--120\relax
\mciteBstWouldAddEndPuncttrue
\mciteSetBstMidEndSepPunct{\mcitedefaultmidpunct}
{\mcitedefaultendpunct}{\mcitedefaultseppunct}\relax
\EndOfBibitem
\bibitem[Hellewell \latin{et~al.}(1994)Hellewell, Bruce, Feinstein, Orringer,
  Williams, and Bowen]{Hellewell1994}
Hellewell,~S.~B.; Bruce,~A.; Feinstein,~G.; Orringer,~J.; Williams,~W.;
  Bowen,~W.~D. {Rat Liver and Kidney Contain High Densities of $\Sigma$1 and
  $\Sigma$2 Receptors: Characterization by Ligand Binding and Photoaffinity
  Labeling}. \emph{Eur. J. Pharmacol. Mol. Pharmacol.} \textbf{1994},
  \emph{268}, 9--18\relax
\mciteBstWouldAddEndPuncttrue
\mciteSetBstMidEndSepPunct{\mcitedefaultmidpunct}
{\mcitedefaultendpunct}{\mcitedefaultseppunct}\relax
\EndOfBibitem
\bibitem[Azzariti \latin{et~al.}(2006)Azzariti, Colabufo, Berardi, Porcelli,
  Niso, Simone, Perrone, and Paradiso]{Azzariti2006}
Azzariti,~A.; Colabufo,~N.~A.; Berardi,~F.; Porcelli,~L.; Niso,~M.;
  Simone,~G.~M.; Perrone,~R.; Paradiso,~A. {Cyclohexylpiperazine Derivative
  PB28, a $\sigma$2 Agonist and $\sigma$1 Antagonist Receptor, Inhibits Cell
  Growth, Modulates P-Glycoprotein, and Synergizes With Anthracyclines in
  Breast Cancer}. \emph{Mol. Cancer Ther.} \textbf{2006}, \emph{5},
  1807--1816\relax
\mciteBstWouldAddEndPuncttrue
\mciteSetBstMidEndSepPunct{\mcitedefaultmidpunct}
{\mcitedefaultendpunct}{\mcitedefaultseppunct}\relax
\EndOfBibitem
\bibitem[Cuzzocrea and Catania(2011)Cuzzocrea, and Catania]{Cuzzocrea2011}
Cuzzocrea,~S.; Catania,~M. {Pharmacological and Clinical Overview of
  Cloperastine in Treatment of Cough}. \emph{Ther. Clin. Risk Manag.}
  \textbf{2011}, 83\relax
\mciteBstWouldAddEndPuncttrue
\mciteSetBstMidEndSepPunct{\mcitedefaultmidpunct}
{\mcitedefaultendpunct}{\mcitedefaultseppunct}\relax
\EndOfBibitem
\bibitem[Phillips \latin{et~al.}(2005)Phillips, Braun, Wang, Gumbart,
  Tajkhorshid, Villa, Chipot, Skeel, Kal{\'{e}}, and Schulten]{Phillips2005}
Phillips,~J.~C.; Braun,~R.; Wang,~W.; Gumbart,~J.; Tajkhorshid,~E.; Villa,~E.;
  Chipot,~C.; Skeel,~R.~D.; Kal{\'{e}},~L.; Schulten,~K. {Scalable Molecular
  Dynamics With Namd}. \emph{J. Comput. Chem.} \textbf{2005}, \emph{26},
  1781--1802\relax
\mciteBstWouldAddEndPuncttrue
\mciteSetBstMidEndSepPunct{\mcitedefaultmidpunct}
{\mcitedefaultendpunct}{\mcitedefaultseppunct}\relax
\EndOfBibitem
\bibitem[Klauda \latin{et~al.}(2010)Klauda, Venable, Freites, Connor, Tobias,
  Mondragon-ramirez, Vorobyov, Mackerell, and Pastor]{Klauda2010}
Klauda,~J.~B.; Venable,~R.~M.; Freites,~J.~A.; Connor,~J. W.~O.; Tobias,~D.~J.;
  Mondragon-ramirez,~C.; Vorobyov,~I.; Mackerell,~A.~D.; Pastor,~R.~W. {Update
  of the Charmm All-Atom Additive Force Field for Lipids: Validation on Six
  Lipid Types}. \emph{J. Phys. Chem. B} \textbf{2010}, \emph{2},
  7830--7843\relax
\mciteBstWouldAddEndPuncttrue
\mciteSetBstMidEndSepPunct{\mcitedefaultmidpunct}
{\mcitedefaultendpunct}{\mcitedefaultseppunct}\relax
\EndOfBibitem
\bibitem[MacKerell \latin{et~al.}(1998)MacKerell, Bashford, Bellott, Dunbrack,
  Evanseck, Field, Fischer, Gao, Guo, Ha, Joseph-McCarthy, Kuchnir, Kuczera,
  Lau, Mattos, Michnick, Ngo, Nguyen, Prodhom, Reiher, Roux, Schlenkrich,
  Smith, Stote, Straub, Watanabe, Wi{\'{o}}rkiewicz-Kuczera, Yin, and
  Karplus]{MacKerell1998}
MacKerell,~A.~D.; Bashford,~D.; Bellott,~M.; Dunbrack,~R.~L.; Evanseck,~J.~D.;
  Field,~M.~J.; Fischer,~S.; Gao,~J.; Guo,~H.; Ha,~S.; Joseph-McCarthy,~D.;
  Kuchnir,~L.; Kuczera,~K.; Lau,~F. T.~K.; Mattos,~C.; Michnick,~S.; Ngo,~T.;
  Nguyen,~D.~T.; Prodhom,~B.; Reiher,~W.~E. \latin{et~al.}  {All-Atom Empirical
  Potential for Molecular Modeling and Dynamics Studies of Proteins}. \emph{J.
  Phys. Chem. B} \textbf{1998}, \emph{102}, 3586--3616\relax
\mciteBstWouldAddEndPuncttrue
\mciteSetBstMidEndSepPunct{\mcitedefaultmidpunct}
{\mcitedefaultendpunct}{\mcitedefaultseppunct}\relax
\EndOfBibitem
\bibitem[Ryckaert \latin{et~al.}(1977)Ryckaert, Ciccotti, and
  Berendsen]{Ryckaert1977}
Ryckaert,~J.~P.; Ciccotti,~G.; Berendsen,~H.~J. {Numerical Integration of the
  Cartesian Equations of Motion of a System With Constraints: Molecular
  Dynamics of N-Alkanes}. \emph{J. Comput. Phys.} \textbf{1977}, \emph{23},
  327--341\relax
\mciteBstWouldAddEndPuncttrue
\mciteSetBstMidEndSepPunct{\mcitedefaultmidpunct}
{\mcitedefaultendpunct}{\mcitedefaultseppunct}\relax
\EndOfBibitem
\bibitem[Darden \latin{et~al.}(1993)Darden, York, and Pedersen]{Darden1993}
Darden,~T.; York,~D.; Pedersen,~L. {Particle Mesh Ewald: an Nlog(n) Method for
  Ewald Sums in Large Systems}. \emph{J. Chem. Phys.} \textbf{1993}, \emph{98},
  10089--10092\relax
\mciteBstWouldAddEndPuncttrue
\mciteSetBstMidEndSepPunct{\mcitedefaultmidpunct}
{\mcitedefaultendpunct}{\mcitedefaultseppunct}\relax
\EndOfBibitem
\bibitem[Loncharich \latin{et~al.}(1992)Loncharich, Brooks, and
  Pastor]{Loncharich1992}
Loncharich,~R.~J.; Brooks,~B.~R.; Pastor,~R.~W. {Langevin Dynamics of Peptides:
  the Frictional Dependence of Isomerization Rates of
  \textit{N}-Acetylalanyl-\textit{N'}-‐Methylamide}. \emph{Biopolymers}
  \textbf{1992}, \emph{32}, 523--535\relax
\mciteBstWouldAddEndPuncttrue
\mciteSetBstMidEndSepPunct{\mcitedefaultmidpunct}
{\mcitedefaultendpunct}{\mcitedefaultseppunct}\relax
\EndOfBibitem
\bibitem[{Martyna, Tobias}(1994)]{MartynaTobiasKlein1994}
{Martyna, Tobias},~K. {Constant Pressure Molecular Dynamics Algorithms}.
  \emph{J. Chem. Phys.} \textbf{1994}, \emph{101}, 4177--4189\relax
\mciteBstWouldAddEndPuncttrue
\mciteSetBstMidEndSepPunct{\mcitedefaultmidpunct}
{\mcitedefaultendpunct}{\mcitedefaultseppunct}\relax
\EndOfBibitem
\bibitem[Feller(1995)]{Feller1995}
Feller,~S.~E. {Constant Pressure Molecular Dynamics Simulation: the Langevin
  Piston Method}. \emph{J. Chem. Phys.} \textbf{1995}, \emph{103},
  4613--4621\relax
\mciteBstWouldAddEndPuncttrue
\mciteSetBstMidEndSepPunct{\mcitedefaultmidpunct}
{\mcitedefaultendpunct}{\mcitedefaultseppunct}\relax
\EndOfBibitem
\bibitem[Horvath and Daum(2013)Horvath, and Daum]{Horvath2013}
Horvath,~S.~E.; Daum,~G. {Lipids of Mitochondria}. \emph{Prog. Lipid Res.}
  \textbf{2013}, \emph{52}, 590--614\relax
\mciteBstWouldAddEndPuncttrue
\mciteSetBstMidEndSepPunct{\mcitedefaultmidpunct}
{\mcitedefaultendpunct}{\mcitedefaultseppunct}\relax
\EndOfBibitem
\bibitem[Jo \latin{et~al.}(2008)Jo, Kim, Iyer, and Im]{Jo2007}
Jo,~S.; Kim,~T.; Iyer,~V.~G.; Im,~W. {Charmm-Gui: a Web-Based Graphical User
  Interface for Charmm}. \emph{J. Comput. Chem.} \textbf{2008}, \emph{29},
  1859--1865\relax
\mciteBstWouldAddEndPuncttrue
\mciteSetBstMidEndSepPunct{\mcitedefaultmidpunct}
{\mcitedefaultendpunct}{\mcitedefaultseppunct}\relax
\EndOfBibitem
\bibitem[Vanommeslaeghe \latin{et~al.}(2010)Vanommeslaeghe, Hatcher, Acharya,
  Kundu, Zhong, Shim, Darian, Guvench, Lopes, Vorobyov, and
  Mackerell]{Vanommeslaeghe2010}
Vanommeslaeghe,~K.; Hatcher,~E.; Acharya,~C.; Kundu,~S.; Zhong,~S.; Shim,~J.;
  Darian,~E.; Guvench,~O.; Lopes,~P.; Vorobyov,~I.; Mackerell,~A.~D. {Charmm
  General Force Field: a Force Field for Drug-Like Molecules Compatible With
  the Charmm All-Atom Additive Biological Force Fields}. \emph{J. Comput.
  Chem.} \textbf{2010}, \emph{31}, 671--690\relax
\mciteBstWouldAddEndPuncttrue
\mciteSetBstMidEndSepPunct{\mcitedefaultmidpunct}
{\mcitedefaultendpunct}{\mcitedefaultseppunct}\relax
\EndOfBibitem
\bibitem[Vanommeslaeghe and MacKerell(2012)Vanommeslaeghe, and
  MacKerell]{Vanommeslaeghe2012}
Vanommeslaeghe,~K.; MacKerell,~A.~D. {Automation of the Charmm General Force
  Field (cgenff) I: Bond Perception and Atom Typing}. \emph{J. Chem. Inf.
  Model.} \textbf{2012}, \emph{52}, 3144--3154\relax
\mciteBstWouldAddEndPuncttrue
\mciteSetBstMidEndSepPunct{\mcitedefaultmidpunct}
{\mcitedefaultendpunct}{\mcitedefaultseppunct}\relax
\EndOfBibitem
\bibitem[Cheng \latin{et~al.}(2007)Cheng, Zhao, Li, Lin, Xu, Zhang, Li, Wang,
  and Lai]{Cheng2007}
Cheng,~T.; Zhao,~Y.; Li,~X.; Lin,~F.; Xu,~Y.; Zhang,~X.; Li,~Y.; Wang,~R.;
  Lai,~L. {Computation of Octanol-Water Partition Coefficients by Guiding an
  Additive Nodel With Knowledge}. \emph{J. Chem. Inf. Model.} \textbf{2007},
  \emph{47}, 2140--2148\relax
\mciteBstWouldAddEndPuncttrue
\mciteSetBstMidEndSepPunct{\mcitedefaultmidpunct}
{\mcitedefaultendpunct}{\mcitedefaultseppunct}\relax
\EndOfBibitem
\end{mcitethebibliography}

\end{document}


\setcounter{figure}{0}

\renewcommand{\thetable}{S\arabic{table}}
\begin{table*}[!ht]
  \centering
  \begin{threeparttable}
  \caption{Lipid composition of membranes\tnote{1}.}
  \label{tab:drug_data}
  \begin{tabular}{lccccccc}
  \toprule
    Membrane & POPC & POPE & POPI & POPS & CL & PSM & Cholesterol \\
  \midrule
    Plasma\tnote{2} & 25 & 15 & 5 & 6 & 1 & 11 & 38 \\
    Lysosome\tnote{2} & 25 & 9 & 3 & 2 & 1 & 13 & 48 \\
    Golgi\tnote{2} & 47 & 19 & 11 & 6 & 1 & 7 & 9 \\
    Mitochondrial\tnote{2} & 43 & 33 & 5 & 1 & 14 & 1 & 4 \\
    \makecell[l]{Endoplasmic \\ Reticulum\tnote{3}} & 72 & 17 & 6 & 4 & 0 & 2 & 7 \\
    \makecell[l]{Coronavirus \\ wo spike protein\tnote{3}}
     & 65 & 10 & 9 & 5 & 0 & 4 & 7 \\
    \makecell[l]{Coronavirus \\ w spike protein\tnote{3}} & 89 & 13 & 13 & 7 & 0 & 5 & 9 \\
  \bottomrule
  \end{tabular}
  \label{tab:membranes_lipid_composition}
\begin{tablenotes}
\small
\item[1] values represent the number of lipid on each leaflet.
\item[2] based on the ratio provided by Horvath~\etal~\textsuperscript{34}.
\item[3] based on the ratio provided by van~Genderen~\etal~\textsuperscript{15}.
\end{tablenotes}
\end{threeparttable}
\end{table*}

\begin{landscape}

\setlength{\tabcolsep}{3pt}
\renewcommand{\arraystretch}{0.5}
\small
\begin{ThreePartTable}
  \begin{longtable}{l|c|ccccccc|ccccccc}
  \caption{Parameters of the LDA model for all the modeled drugs} \\
  \toprule
    Drug & logP & $P_p$\tnote{3} & $P_l$\tnote{4} & $P_g$\tnote{5} & $P_m$\tnote{6} & $P_e$\tnote{7} & $P_{cs}$\tnote{8} & $P_c$\tnote{9} & $\tau_p$\tnote{3} & $\tau_l$\tnote{4} & $\tau_g$\tnote{5} & $\tau_m$\tnote{6} & $\tau_e$\tnote{7} & $\tau_{cs}$\tnote{8} & $\tau_c$\tnote{9} \\
  \midrule
4E2RCat & 5.40\tnote{2} & \makecell{3.1$\pm$\\0.5\%} & \makecell{2.3$\pm$\\0.3\%} & \makecell{8.6$\pm$\\0.3\%} & \makecell{11.2$\pm$\\0.4\%} & \makecell{10.7$\pm$\\0.3\%} & \makecell{10.5$\pm$\\0.5\%} & \makecell{10.6$\pm$\\0.3\%} & \makecell{50.1$\pm$\\\ns{4.1}} & \makecell{56.3$\pm$\\\ns{2.6}} & \makecell{24.0$\pm$\\\ns{0.9}} & \makecell{21.2$\pm$\\\ns{0.7}} & \makecell{16.7$\pm$\\\ns{0.8}} & \makecell{25.2$\pm$\\\ns{2.3}} & \makecell{18.7$\pm$\\\ns{0.5}}\\ 
ABBV-744 & 3.90\tnote{2} & \makecell{4.3$\pm$\\0.6\%} & \makecell{3.3$\pm$\\0.4\%} & \makecell{11.6$\pm$\\0.4\%} & \makecell{14.8$\pm$\\0.5\%} & \makecell{14.1$\pm$\\0.4\%} & \makecell{13.5$\pm$\\0.5\%} & \makecell{14.2$\pm$\\0.4\%} & \makecell{36.8$\pm$\\\ns{2.6}} & \makecell{39.1$\pm$\\\ns{1.7}} & \makecell{17.6$\pm$\\\ns{0.7}} & \makecell{15.7$\pm$\\\ns{0.5}} & \makecell{12.5$\pm$\\\ns{0.6}} & \makecell{19.2$\pm$\\\ns{1.8}} & \makecell{13.7$\pm$\\\ns{0.3}}\\ 
AC-55541 & 4.10\tnote{2} & \makecell{3.4$\pm$\\0.5\%} & \makecell{2.7$\pm$\\0.3\%} & \makecell{8.9$\pm$\\0.3\%} & \makecell{11.6$\pm$\\0.4\%} & \makecell{10.9$\pm$\\0.3\%} & \makecell{10.7$\pm$\\0.5\%} & \makecell{10.9$\pm$\\0.3\%} & \makecell{46.1$\pm$\\\ns{3.2}} & \makecell{49.3$\pm$\\\ns{2.1}} & \makecell{23.2$\pm$\\\ns{0.9}} & \makecell{20.4$\pm$\\\ns{0.6}} & \makecell{16.4$\pm$\\\ns{0.8}} & \makecell{24.6$\pm$\\\ns{2.3}} & \makecell{18.1$\pm$\\\ns{0.5}}\\ 
AZ3451 & 6.90\tnote{2} & \makecell{4.9$\pm$\\0.4\%} & \makecell{4.1$\pm$\\0.3\%} & \makecell{10.0$\pm$\\0.3\%} & \makecell{11.8$\pm$\\0.3\%} & \makecell{11.7$\pm$\\0.2\%} & \makecell{11.0$\pm$\\0.4\%} & \makecell{11.9$\pm$\\0.3\%} & \makecell{32.1$\pm$\\\ns{2.0}} & \makecell{31.9$\pm$\\\ns{1.0}} & \makecell{20.6$\pm$\\\ns{0.9}} & \makecell{20.0$\pm$\\\ns{0.7}} & \makecell{15.2$\pm$\\\ns{0.6}} & \makecell{24.0$\pm$\\\ns{2.4}} & \makecell{16.6$\pm$\\\ns{0.5}}\\ 
AZ8838 & 1.80\tnote{2} & \makecell{6.1$\pm$\\0.6\%} & \makecell{5.2$\pm$\\0.5\%} & \makecell{13.9$\pm$\\0.5\%} & \makecell{16.9$\pm$\\0.7\%} & \makecell{16.5$\pm$\\0.4\%} & \makecell{17.0$\pm$\\0.7\%} & \makecell{17.1$\pm$\\0.5\%} & \makecell{25.7$\pm$\\\ns{1.3}} & \makecell{25.2$\pm$\\\ns{0.8}} & \makecell{14.6$\pm$\\\ns{0.5}} & \makecell{13.8$\pm$\\\ns{0.4}} & \makecell{10.7$\pm$\\\ns{0.4}} & \makecell{15.1$\pm$\\\ns{1.4}} & \makecell{11.4$\pm$\\\ns{0.3}}\\ 
Apicidin & 4.40\tnote{2} & \makecell{2.7$\pm$\\0.3\%} & \makecell{2.3$\pm$\\0.2\%} & \makecell{6.2$\pm$\\0.2\%} & \makecell{8.0$\pm$\\0.3\%} & \makecell{7.7$\pm$\\0.2\%} & \makecell{8.0$\pm$\\0.4\%} & \makecell{7.8$\pm$\\0.3\%} & \makecell{58.1$\pm$\\\ns{3.0}} & \makecell{56.6$\pm$\\\ns{1.8}} & \makecell{33.6$\pm$\\\ns{1.2}} & \makecell{30.5$\pm$\\\ns{1.0}} & \makecell{23.7$\pm$\\\ns{1.1}} & \makecell{33.6$\pm$\\\ns{2.9}} & \makecell{25.7$\pm$\\\ns{0.6}}\\ 
Atovaquone & 5.80\tnote{1} & \makecell{5.7$\pm$\\0.5\%} & \makecell{4.7$\pm$\\0.3\%} & \makecell{11.2$\pm$\\0.2\%} & \makecell{12.9$\pm$\\0.2\%} & \makecell{13.0$\pm$\\0.2\%} & \makecell{12.3$\pm$\\0.4\%} & \makecell{13.0$\pm$\\0.3\%} & \makecell{27.4$\pm$\\\ns{1.9}} & \makecell{28.0$\pm$\\\ns{0.8}} & \makecell{18.2$\pm$\\\ns{0.9}} & \makecell{18.3$\pm$\\\ns{0.7}} & \makecell{13.6$\pm$\\\ns{0.5}} & \makecell{21.3$\pm$\\\ns{2.2}} & \makecell{15.0$\pm$\\\ns{0.6}}\\ 
Azacitidine & -2.17\tnote{1} & \makecell{5.3$\pm$\\0.7\%} & \makecell{4.2$\pm$\\0.4\%} & \makecell{13.2$\pm$\\0.5\%} & \makecell{17.2$\pm$\\0.7\%} & \makecell{16.6$\pm$\\0.5\%} & \makecell{15.8$\pm$\\0.7\%} & \makecell{16.3$\pm$\\0.5\%} & \makecell{4.4$\pm$\\\us{0.3}} & \makecell{4.6$\pm$\\\us{0.2}} & \makecell{2.4$\pm$\\\us{0.1}} & \makecell{2.2$\pm$\\\us{0.1}} & \makecell{1.7$\pm$\\\us{0.1}} & \makecell{2.7$\pm$\\\us{0.3}} & \makecell{1.9$\pm$\\\us{0.1}}\\ 
Bafilomycin A1 & 6.00\tnote{2} & \makecell{1.6$\pm$\\0.3\%} & \makecell{1.3$\pm$\\0.2\%} & \makecell{6.3$\pm$\\0.3\%} & \makecell{8.3$\pm$\\0.4\%} & \makecell{7.8$\pm$\\0.3\%} & \makecell{8.5$\pm$\\0.4\%} & \makecell{8.3$\pm$\\0.3\%} & \makecell{95.8$\pm$\\\ns{13.0}} & \makecell{104.6$\pm$\\\ns{8.2}} & \makecell{33.2$\pm$\\\ns{1.2}} & \makecell{29.1$\pm$\\\ns{1.0}} & \makecell{23.1$\pm$\\\ns{1.2}} & \makecell{31.7$\pm$\\\ns{2.7}} & \makecell{24.1$\pm$\\\ns{0.3}}\\ 
CB5083 & 3.10\tnote{2} & \makecell{3.5$\pm$\\0.4\%} & \makecell{3.0$\pm$\\0.3\%} & \makecell{9.8$\pm$\\0.3\%} & \makecell{12.1$\pm$\\0.5\%} & \makecell{11.7$\pm$\\0.3\%} & \makecell{11.9$\pm$\\0.5\%} & \makecell{12.2$\pm$\\0.4\%} & \makecell{44.6$\pm$\\\ns{2.6}} & \makecell{43.5$\pm$\\\ns{1.7}} & \makecell{20.9$\pm$\\\ns{0.8}} & \makecell{19.6$\pm$\\\ns{0.6}} & \makecell{15.2$\pm$\\\ns{0.6}} & \makecell{22.1$\pm$\\\ns{1.9}} & \makecell{16.1$\pm$\\\ns{0.4}}\\ 
CCT 365623 & 2.49\tnote{2} & \makecell{4.0$\pm$\\0.5\%} & \makecell{3.5$\pm$\\0.3\%} & \makecell{10.5$\pm$\\0.3\%} & \makecell{13.0$\pm$\\0.5\%} & \makecell{12.7$\pm$\\0.3\%} & \makecell{12.5$\pm$\\0.6\%} & \makecell{13.0$\pm$\\0.4\%} & \makecell{39.0$\pm$\\\ns{2.0}} & \makecell{37.7$\pm$\\\ns{1.3}} & \makecell{19.5$\pm$\\\ns{0.7}} & \makecell{18.1$\pm$\\\ns{0.5}} & \makecell{14.1$\pm$\\\ns{0.6}} & \makecell{20.9$\pm$\\\ns{1.8}} & \makecell{15.0$\pm$\\\ns{0.4}}\\ 
CPI-0610 & 3.00\tnote{2} & \makecell{5.9$\pm$\\0.5\%} & \makecell{4.8$\pm$\\0.4\%} & \makecell{12.3$\pm$\\0.3\%} & \makecell{15.1$\pm$\\0.5\%} & \makecell{14.8$\pm$\\0.4\%} & \makecell{14.4$\pm$\\0.6\%} & \makecell{14.8$\pm$\\0.4\%} & \makecell{26.2$\pm$\\\ns{1.6}} & \makecell{26.9$\pm$\\\ns{0.8}} & \makecell{16.5$\pm$\\\ns{0.6}} & \makecell{15.4$\pm$\\\ns{0.5}} & \makecell{11.9$\pm$\\\ns{0.5}} & \makecell{17.9$\pm$\\\ns{1.6}} & \makecell{13.1$\pm$\\\ns{0.3}}\\ 
Camostat & 1.10\tnote{2} & \makecell{4.1$\pm$\\0.4\%} & \makecell{3.6$\pm$\\0.3\%} & \makecell{9.1$\pm$\\0.3\%} & \makecell{10.8$\pm$\\0.3\%} & \makecell{10.7$\pm$\\0.2\%} & \makecell{10.4$\pm$\\0.4\%} & \makecell{11.1$\pm$\\0.3\%} & \makecell{41.0$\pm$\\\ns{2.1}} & \makecell{39.1$\pm$\\\ns{1.1}} & \makecell{24.5$\pm$\\\ns{1.2}} & \makecell{24.0$\pm$\\\ns{0.8}} & \makecell{18.1$\pm$\\\ns{0.6}} & \makecell{27.5$\pm$\\\ns{2.8}} & \makecell{19.3$\pm$\\\ns{0.6}}\\ 
Captopril & 0.34\tnote{1} & \makecell{4.3$\pm$\\0.7\%} & \makecell{3.5$\pm$\\0.4\%} & \makecell{13.4$\pm$\\0.5\%} & \makecell{17.3$\pm$\\0.8\%} & \makecell{16.4$\pm$\\0.5\%} & \makecell{16.3$\pm$\\0.7\%} & \makecell{16.9$\pm$\\0.6\%} & \makecell{52.9$\pm$\\\ns{4.7}} & \makecell{54.7$\pm$\\\ns{3.0}} & \makecell{22.4$\pm$\\\ns{0.8}} & \makecell{20.0$\pm$\\\ns{0.6}} & \makecell{15.8$\pm$\\\ns{0.8}} & \makecell{23.5$\pm$\\\ns{2.2}} & \makecell{17.0$\pm$\\\ns{0.3}}\\ 
Chloramphenicol & 1.14\tnote{1} & \makecell{4.7$\pm$\\0.4\%} & \makecell{4.1$\pm$\\0.4\%} & \makecell{11.3$\pm$\\0.3\%} & \makecell{14.0$\pm$\\0.5\%} & \makecell{13.7$\pm$\\0.3\%} & \makecell{13.0$\pm$\\0.7\%} & \makecell{13.9$\pm$\\0.4\%} & \makecell{36.0$\pm$\\\ns{1.7}} & \makecell{34.0$\pm$\\\ns{1.0}} & \makecell{19.4$\pm$\\\ns{0.8}} & \makecell{18.0$\pm$\\\ns{0.5}} & \makecell{13.8$\pm$\\\ns{0.5}} & \makecell{21.6$\pm$\\\ns{1.8}} & \makecell{15.1$\pm$\\\ns{0.4}}\\ 
Chloroquine & 4.63\tnote{1} & \makecell{3.2$\pm$\\0.5\%} & \makecell{2.7$\pm$\\0.3\%} & \makecell{9.7$\pm$\\0.4\%} & \makecell{12.5$\pm$\\0.6\%} & \makecell{12.0$\pm$\\0.4\%} & \makecell{12.1$\pm$\\0.6\%} & \makecell{12.3$\pm$\\0.4\%} & \makecell{48.4$\pm$\\\ns{3.6}} & \makecell{48.2$\pm$\\\ns{2.2}} & \makecell{21.2$\pm$\\\ns{0.7}} & \makecell{18.8$\pm$\\\ns{0.6}} & \makecell{14.8$\pm$\\\ns{0.7}} & \makecell{21.6$\pm$\\\ns{1.8}} & \makecell{16.0$\pm$\\\ns{0.3}}\\ 
Clemastine & 5.00\tnote{2} & \makecell{5.9$\pm$\\0.5\%} & \makecell{5.2$\pm$\\0.4\%} & \makecell{13.1$\pm$\\0.3\%} & \makecell{15.3$\pm$\\0.4\%} & \makecell{15.2$\pm$\\0.3\%} & \makecell{14.9$\pm$\\0.6\%} & \makecell{15.6$\pm$\\0.4\%} & \makecell{26.2$\pm$\\\ns{1.3}} & \makecell{25.2$\pm$\\\ns{0.7}} & \makecell{15.4$\pm$\\\ns{0.7}} & \makecell{15.1$\pm$\\\ns{0.5}} & \makecell{11.5$\pm$\\\ns{0.4}} & \makecell{17.2$\pm$\\\ns{1.6}} & \makecell{12.4$\pm$\\\ns{0.4}}\\ 
Cloperastine & 4.80\tnote{2} & \makecell{5.3$\pm$\\0.5\%} & \makecell{4.5$\pm$\\0.4\%} & \makecell{12.5$\pm$\\0.3\%} & \makecell{14.7$\pm$\\0.5\%} & \makecell{14.5$\pm$\\0.3\%} & \makecell{14.2$\pm$\\0.6\%} & \makecell{15.1$\pm$\\0.4\%} & \makecell{29.5$\pm$\\\ns{1.5}} & \makecell{28.9$\pm$\\\ns{0.9}} & \makecell{16.1$\pm$\\\ns{0.7}} & \makecell{15.9$\pm$\\\ns{0.5}} & \makecell{12.2$\pm$\\\ns{0.4}} & \makecell{18.2$\pm$\\\ns{1.7}} & \makecell{12.8$\pm$\\\ns{0.4}}\\ 
Compound 10 & 4.07\tnote{2} & \makecell{2.4$\pm$\\0.3\%} & \makecell{2.1$\pm$\\0.2\%} & \makecell{7.7$\pm$\\0.3\%} & \makecell{9.8$\pm$\\0.4\%} & \makecell{9.5$\pm$\\0.3\%} & \makecell{8.8$\pm$\\0.4\%} & \makecell{9.9$\pm$\\0.3\%} & \makecell{65.4$\pm$\\\ns{4.3}} & \makecell{63.9$\pm$\\\ns{2.8}} & \makecell{26.9$\pm$\\\ns{1.1}} & \makecell{24.5$\pm$\\\ns{0.8}} & \makecell{18.9$\pm$\\\ns{0.8}} & \makecell{30.3$\pm$\\\ns{2.7}} & \makecell{20.2$\pm$\\\ns{0.5}}\\ 
Compound 2 & 2.47\tnote{2} & \makecell{2.3$\pm$\\0.3\%} & \makecell{1.9$\pm$\\0.2\%} & \makecell{6.1$\pm$\\0.3\%} & \makecell{7.9$\pm$\\0.4\%} & \makecell{7.4$\pm$\\0.2\%} & \makecell{7.6$\pm$\\0.5\%} & \makecell{7.7$\pm$\\0.3\%} & \makecell{68.9$\pm$\\\ns{4.4}} & \makecell{70.4$\pm$\\\ns{2.5}} & \makecell{34.7$\pm$\\\ns{1.1}} & \makecell{30.8$\pm$\\\ns{0.9}} & \makecell{24.5$\pm$\\\ns{1.2}} & \makecell{35.7$\pm$\\\ns{2.6}} & \makecell{26.2$\pm$\\\ns{0.5}}\\ 
DBeQ & 5.30\tnote{2} & \makecell{4.8$\pm$\\0.5\%} & \makecell{4.3$\pm$\\0.4\%} & \makecell{11.8$\pm$\\0.3\%} & \makecell{14.1$\pm$\\0.5\%} & \makecell{13.8$\pm$\\0.3\%} & \makecell{13.8$\pm$\\0.6\%} & \makecell{14.3$\pm$\\0.4\%} & \makecell{32.4$\pm$\\\ns{1.4}} & \makecell{30.5$\pm$\\\ns{0.9}} & \makecell{17.3$\pm$\\\ns{0.7}} & \makecell{16.5$\pm$\\\ns{0.5}} & \makecell{12.8$\pm$\\\ns{0.5}} & \makecell{18.7$\pm$\\\ns{1.7}} & \makecell{13.6$\pm$\\\ns{0.4}}\\ 
Dabrafenib & 4.80\tnote{2} & \makecell{2.3$\pm$\\0.4\%} & \makecell{1.8$\pm$\\0.2\%} & \makecell{7.9$\pm$\\0.4\%} & \makecell{10.7$\pm$\\0.5\%} & \makecell{10.0$\pm$\\0.4\%} & \makecell{9.5$\pm$\\0.6\%} & \makecell{10.2$\pm$\\0.4\%} & \makecell{68.0$\pm$\\\ns{6.4}} & \makecell{71.7$\pm$\\\ns{4.0}} & \makecell{26.4$\pm$\\\ns{1.0}} & \makecell{22.4$\pm$\\\ns{0.7}} & \makecell{18.0$\pm$\\\ns{1.0}} & \makecell{27.9$\pm$\\\ns{2.1}} & \makecell{19.5$\pm$\\\ns{0.4}}\\ 
Daunorubicin & 1.83\tnote{1} & \makecell{1.4$\pm$\\0.3\%} & \makecell{1.1$\pm$\\0.2\%} & \makecell{5.4$\pm$\\0.3\%} & \makecell{7.6$\pm$\\0.4\%} & \makecell{7.2$\pm$\\0.2\%} & \makecell{7.3$\pm$\\0.5\%} & \makecell{7.2$\pm$\\0.3\%} & \makecell{116.1$\pm$\\\ns{16.1}} & \makecell{124.0$\pm$\\\ns{9.2}} & \makecell{39.8$\pm$\\\ns{1.4}} & \makecell{32.6$\pm$\\\ns{1.0}} & \makecell{25.6$\pm$\\\ns{1.2}} & \makecell{37.8$\pm$\\\ns{2.7}} & \makecell{28.5$\pm$\\\ns{0.4}}\\ 
E-52862 & 3.50\tnote{2} & \makecell{3.7$\pm$\\0.4\%} & \makecell{3.3$\pm$\\0.3\%} & \makecell{10.1$\pm$\\0.3\%} & \makecell{12.3$\pm$\\0.4\%} & \makecell{12.0$\pm$\\0.3\%} & \makecell{12.2$\pm$\\0.5\%} & \makecell{12.5$\pm$\\0.4\%} & \makecell{41.9$\pm$\\\ns{2.1}} & \makecell{40.1$\pm$\\\ns{1.4}} & \makecell{20.3$\pm$\\\ns{0.8}} & \makecell{19.3$\pm$\\\ns{0.6}} & \makecell{14.8$\pm$\\\ns{0.6}} & \makecell{21.5$\pm$\\\ns{1.9}} & \makecell{15.7$\pm$\\\ns{0.4}}\\ 
Entacapone & 2.10\tnote{2} & \makecell{2.8$\pm$\\0.4\%} & \makecell{2.1$\pm$\\0.3\%} & \makecell{8.3$\pm$\\0.4\%} & \makecell{11.5$\pm$\\0.5\%} & \makecell{10.7$\pm$\\0.4\%} & \makecell{10.7$\pm$\\0.6\%} & \makecell{10.8$\pm$\\0.4\%} & \makecell{57.5$\pm$\\\ns{5.5}} & \makecell{63.8$\pm$\\\ns{3.5}} & \makecell{25.1$\pm$\\\ns{0.9}} & \makecell{20.8$\pm$\\\ns{0.6}} & \makecell{16.8$\pm$\\\ns{0.9}} & \makecell{25.0$\pm$\\\ns{1.8}} & \makecell{18.6$\pm$\\\ns{0.3}}\\ 
FK-506 & 3.80\tnote{2} & \makecell{0.4$\pm$\\0.1\%} & \makecell{0.2$\pm$\\0.1\%} & \makecell{2.4$\pm$\\0.2\%} & \makecell{4.1$\pm$\\0.3\%} & \makecell{3.5$\pm$\\0.3\%} & \makecell{3.8$\pm$\\0.5\%} & \makecell{3.6$\pm$\\0.3\%} & \makecell{384.9$\pm$\\\ns{121.9}} & \makecell{634.7$\pm$\\\ns{116.6}} & \makecell{88.0$\pm$\\\ns{5.2}} & \makecell{60.3$\pm$\\\ns{2.7}} & \makecell{52.3$\pm$\\\ns{4.7}} & \makecell{72.2$\pm$\\\ns{3.3}} & \makecell{57.7$\pm$\\\ns{1.7}}\\ 
Favipiravir & -0.60\tnote{2} & \makecell{8.5$\pm$\\1.1\%} & \makecell{6.5$\pm$\\0.7\%} & \makecell{21.1$\pm$\\0.6\%} & \makecell{26.4$\pm$\\0.7\%} & \makecell{25.3$\pm$\\0.7\%} & \makecell{24.9$\pm$\\0.8\%} & \makecell{25.4$\pm$\\0.7\%} & \makecell{92.8$\pm$\\\ns{6.5}} & \makecell{100.8$\pm$\\\ns{4.0}} & \makecell{50.1$\pm$\\\ns{2.4}} & \makecell{46.6$\pm$\\\ns{1.7}} & \makecell{36.2$\pm$\\\ns{1.6}} & \makecell{54.9$\pm$\\\ns{6.3}} & \makecell{40.1$\pm$\\\ns{1.3}}\\ 
GB110 & 4.30\tnote{2} & \makecell{1.3$\pm$\\0.2\%} & \makecell{1.0$\pm$\\0.1\%} & \makecell{4.3$\pm$\\0.2\%} & \makecell{5.9$\pm$\\0.3\%} & \makecell{5.6$\pm$\\0.2\%} & \makecell{5.6$\pm$\\0.4\%} & \makecell{5.6$\pm$\\0.2\%} & \makecell{125.5$\pm$\\\ns{11.1}} & \makecell{125.6$\pm$\\\ns{6.2}} & \makecell{49.6$\pm$\\\ns{1.6}} & \makecell{41.9$\pm$\\\ns{1.3}} & \makecell{32.7$\pm$\\\ns{1.6}} & \makecell{48.5$\pm$\\\ns{3.4}} & \makecell{36.1$\pm$\\\ns{0.5}}\\ 
H-89 & 3.50\tnote{2} & \makecell{3.7$\pm$\\0.5\%} & \makecell{3.0$\pm$\\0.3\%} & \makecell{9.9$\pm$\\0.4\%} & \makecell{12.5$\pm$\\0.6\%} & \makecell{12.0$\pm$\\0.4\%} & \makecell{12.5$\pm$\\0.6\%} & \makecell{12.5$\pm$\\0.4\%} & \makecell{42.5$\pm$\\\ns{2.8}} & \makecell{43.8$\pm$\\\ns{1.7}} & \makecell{20.7$\pm$\\\ns{0.7}} & \makecell{18.9$\pm$\\\ns{0.6}} & \makecell{14.9$\pm$\\\ns{0.7}} & \makecell{20.9$\pm$\\\ns{1.8}} & \makecell{15.7$\pm$\\\ns{0.3}}\\ 
Haloperidol & 4.30\tnote{1} & \makecell{4.3$\pm$\\0.4\%} & \makecell{3.8$\pm$\\0.3\%} & \makecell{10.1$\pm$\\0.3\%} & \makecell{11.9$\pm$\\0.4\%} & \makecell{11.8$\pm$\\0.2\%} & \makecell{12.0$\pm$\\0.5\%} & \makecell{12.4$\pm$\\0.3\%} & \makecell{36.2$\pm$\\\ns{1.7}} & \makecell{34.1$\pm$\\\ns{1.0}} & \makecell{20.4$\pm$\\\ns{0.9}} & \makecell{19.8$\pm$\\\ns{0.6}} & \makecell{15.1$\pm$\\\ns{0.5}} & \makecell{21.8$\pm$\\\ns{2.1}} & \makecell{15.9$\pm$\\\ns{0.4}}\\ 
IHVR-19029 & 1.67\tnote{1} & \makecell{1.8$\pm$\\0.3\%} & \makecell{1.4$\pm$\\0.2\%} & \makecell{6.5$\pm$\\0.3\%} & \makecell{8.5$\pm$\\0.4\%} & \makecell{8.1$\pm$\\0.3\%} & \makecell{7.8$\pm$\\0.4\%} & \makecell{8.4$\pm$\\0.3\%} & \makecell{91.9$\pm$\\\ns{9.6}} & \makecell{97.8$\pm$\\\ns{6.1}} & \makecell{33.0$\pm$\\\ns{1.2}} & \makecell{29.2$\pm$\\\ns{0.9}} & \makecell{22.8$\pm$\\\ns{1.0}} & \makecell{35.4$\pm$\\\ns{2.9}} & \makecell{24.4$\pm$\\\ns{0.5}}\\ 
Indomethacin & 4.27\tnote{1} & \makecell{6.4$\pm$\\0.5\%} & \makecell{5.6$\pm$\\0.5\%} & \makecell{13.4$\pm$\\0.3\%} & \makecell{15.7$\pm$\\0.5\%} & \makecell{15.6$\pm$\\0.3\%} & \makecell{15.7$\pm$\\0.6\%} & \makecell{16.1$\pm$\\0.4\%} & \makecell{24.4$\pm$\\\ns{1.2}} & \makecell{23.1$\pm$\\\ns{0.6}} & \makecell{15.0$\pm$\\\ns{0.6}} & \makecell{14.7$\pm$\\\ns{0.5}} & \makecell{11.2$\pm$\\\ns{0.4}} & \makecell{16.2$\pm$\\\ns{1.6}} & \makecell{12.0$\pm$\\\ns{0.3}}\\ 
JQ1 & 4.90\tnote{2} & \makecell{4.6$\pm$\\0.5\%} & \makecell{3.7$\pm$\\0.3\%} & \makecell{10.5$\pm$\\0.3\%} & \makecell{12.8$\pm$\\0.5\%} & \makecell{12.4$\pm$\\0.3\%} & \makecell{11.8$\pm$\\0.5\%} & \makecell{12.7$\pm$\\0.4\%} & \makecell{34.0$\pm$\\\ns{2.1}} & \makecell{35.3$\pm$\\\ns{1.1}} & \makecell{19.5$\pm$\\\ns{0.8}} & \makecell{18.4$\pm$\\\ns{0.5}} & \makecell{14.4$\pm$\\\ns{0.6}} & \makecell{22.2$\pm$\\\ns{1.9}} & \makecell{15.4$\pm$\\\ns{0.4}}\\ 
Linezolid & 0.70\tnote{2} & \makecell{3.4$\pm$\\0.4\%} & \makecell{2.9$\pm$\\0.3\%} & \makecell{9.3$\pm$\\0.3\%} & \makecell{11.8$\pm$\\0.4\%} & \makecell{11.5$\pm$\\0.3\%} & \makecell{11.2$\pm$\\0.6\%} & \makecell{11.7$\pm$\\0.4\%} & \makecell{55.5$\pm$\\\ns{3.1}} & \makecell{54.5$\pm$\\\ns{2.0}} & \makecell{26.7$\pm$\\\ns{1.0}} & \makecell{24.4$\pm$\\\ns{0.8}} & \makecell{18.9$\pm$\\\ns{0.8}} & \makecell{28.5$\pm$\\\ns{2.4}} & \makecell{20.4$\pm$\\\ns{0.5}}\\ 
Lisinopril & -1.22\tnote{1} & \makecell{5.4$\pm$\\0.5\%} & \makecell{4.8$\pm$\\0.4\%} & \makecell{12.4$\pm$\\0.3\%} & \makecell{14.7$\pm$\\0.4\%} & \makecell{14.5$\pm$\\0.3\%} & \makecell{14.0$\pm$\\0.5\%} & \makecell{15.0$\pm$\\0.4\%} & \makecell{521.8$\pm$\\\ns{22.9}} & \makecell{486.8$\pm$\\\ns{14.4}} & \makecell{305.8$\pm$\\\ns{14.6}} & \makecell{304.0$\pm$\\\ns{10.8}} & \makecell{227.8$\pm$\\\ns{8.4}} & \makecell{353.6$\pm$\\\ns{37.7}} & \makecell{244.5$\pm$\\\ns{8.9}}\\ 
Loratadine & 5.20\tnote{1} & \makecell{4.0$\pm$\\0.4\%} & \makecell{3.5$\pm$\\0.3\%} & \makecell{10.3$\pm$\\0.3\%} & \makecell{12.5$\pm$\\0.4\%} & \makecell{12.2$\pm$\\0.3\%} & \makecell{12.0$\pm$\\0.5\%} & \makecell{12.7$\pm$\\0.4\%} & \makecell{38.7$\pm$\\\ns{1.9}} & \makecell{37.0$\pm$\\\ns{1.2}} & \makecell{19.9$\pm$\\\ns{0.8}} & \makecell{19.0$\pm$\\\ns{0.6}} & \makecell{14.5$\pm$\\\ns{0.6}} & \makecell{21.7$\pm$\\\ns{1.9}} & \makecell{15.5$\pm$\\\ns{0.4}}\\ 
ML240 & 4.50\tnote{2} & \makecell{3.7$\pm$\\0.5\%} & \makecell{3.0$\pm$\\0.3\%} & \makecell{9.9$\pm$\\0.3\%} & \makecell{12.5$\pm$\\0.5\%} & \makecell{12.0$\pm$\\0.4\%} & \makecell{11.8$\pm$\\0.5\%} & \makecell{12.2$\pm$\\0.4\%} & \makecell{42.6$\pm$\\\ns{2.6}} & \makecell{43.2$\pm$\\\ns{1.7}} & \makecell{20.8$\pm$\\\ns{0.8}} & \makecell{18.9$\pm$\\\ns{0.6}} & \makecell{14.9$\pm$\\\ns{0.7}} & \makecell{22.2$\pm$\\\ns{1.9}} & \makecell{16.1$\pm$\\\ns{0.4}}\\ 
MZ1 & 5.00\tnote{2} & \makecell{2.0$\pm$\\0.3\%} & \makecell{1.6$\pm$\\0.2\%} & \makecell{6.2$\pm$\\0.3\%} & \makecell{7.9$\pm$\\0.4\%} & \makecell{7.5$\pm$\\0.3\%} & \makecell{7.6$\pm$\\0.4\%} & \makecell{7.9$\pm$\\0.3\%} & \makecell{79.4$\pm$\\\ns{6.1}} & \makecell{80.1$\pm$\\\ns{3.9}} & \makecell{34.1$\pm$\\\ns{1.2}} & \makecell{30.9$\pm$\\\ns{1.0}} & \makecell{24.1$\pm$\\\ns{1.2}} & \makecell{35.5$\pm$\\\ns{3.0}} & \makecell{25.5$\pm$\\\ns{0.5}}\\ 
Melperone & 3.30\tnote{2} & \makecell{3.8$\pm$\\0.4\%} & \makecell{3.2$\pm$\\0.3\%} & \makecell{9.9$\pm$\\0.3\%} & \makecell{12.1$\pm$\\0.4\%} & \makecell{11.9$\pm$\\0.3\%} & \makecell{11.5$\pm$\\0.6\%} & \makecell{12.2$\pm$\\0.3\%} & \makecell{41.4$\pm$\\\ns{2.2}} & \makecell{41.0$\pm$\\\ns{1.3}} & \makecell{20.7$\pm$\\\ns{0.8}} & \makecell{19.6$\pm$\\\ns{0.6}} & \makecell{14.9$\pm$\\\ns{0.6}} & \makecell{22.8$\pm$\\\ns{1.9}} & \makecell{16.0$\pm$\\\ns{0.4}}\\ 
Merimepodib & 2.10\tnote{2} & \makecell{3.2$\pm$\\0.3\%} & \makecell{2.8$\pm$\\0.3\%} & \makecell{8.1$\pm$\\0.3\%} & \makecell{9.9$\pm$\\0.4\%} & \makecell{9.7$\pm$\\0.2\%} & \makecell{9.7$\pm$\\0.4\%} & \makecell{10.0$\pm$\\0.3\%} & \makecell{48.9$\pm$\\\ns{2.5}} & \makecell{47.2$\pm$\\\ns{1.6}} & \makecell{26.0$\pm$\\\ns{1.1}} & \makecell{24.4$\pm$\\\ns{0.8}} & \makecell{18.7$\pm$\\\ns{0.7}} & \makecell{27.7$\pm$\\\ns{2.7}} & \makecell{20.1$\pm$\\\ns{0.5}}\\ 
Metformin & -1.30\tnote{2} & \makecell{6.9$\pm$\\0.8\%} & \makecell{5.8$\pm$\\0.6\%} & \makecell{19.1$\pm$\\0.7\%} & \makecell{22.3$\pm$\\0.8\%} & \makecell{22.1$\pm$\\0.4\%} & \makecell{22.0$\pm$\\0.7\%} & \makecell{23.7$\pm$\\0.6\%} & \makecell{481.8$\pm$\\\ns{28.0}} & \makecell{477.9$\pm$\\\ns{18.6}} & \makecell{236.6$\pm$\\\ns{11.6}} & \makecell{238.3$\pm$\\\ns{8.3}} & \makecell{178.0$\pm$\\\ns{6.6}} & \makecell{267.8$\pm$\\\ns{31.0}} & \makecell{184.2$\pm$\\\ns{6.1}}\\ 
Midostaurin & 4.80\tnote{2} & \makecell{6.5$\pm$\\0.6\%} & \makecell{5.8$\pm$\\0.5\%} & \makecell{14.9$\pm$\\0.3\%} & \makecell{17.0$\pm$\\0.5\%} & \makecell{16.8$\pm$\\0.3\%} & \makecell{17.3$\pm$\\0.6\%} & \makecell{17.5$\pm$\\0.4\%} & \makecell{23.7$\pm$\\\ns{1.1}} & \makecell{22.4$\pm$\\\ns{0.7}} & \makecell{13.4$\pm$\\\ns{0.6}} & \makecell{13.5$\pm$\\\ns{0.4}} & \makecell{10.3$\pm$\\\ns{0.3}} & \makecell{14.5$\pm$\\\ns{1.5}} & \makecell{10.9$\pm$\\\ns{0.3}}\\ 
Migalastat & -2.30\tnote{2} & \makecell{7.1$\pm$\\0.8\%} & \makecell{6.1$\pm$\\0.6\%} & \makecell{17.8$\pm$\\0.5\%} & \makecell{21.0$\pm$\\0.8\%} & \makecell{20.5$\pm$\\0.5\%} & \makecell{22.1$\pm$\\0.7\%} & \makecell{21.7$\pm$\\0.6\%} & \makecell{4.5$\pm$\\\us{0.2}} & \makecell{4.3$\pm$\\\us{0.2}} & \makecell{2.4$\pm$\\\us{0.1}} & \makecell{2.4$\pm$\\\us{0.1}} & \makecell{1.8$\pm$\\\us{0.1}} & \makecell{2.6$\pm$\\\us{0.3}} & \makecell{1.9$\pm$\\\us{0.1}}\\ 
Minoxidil & 1.24\tnote{1} & \makecell{4.1$\pm$\\0.7\%} & \makecell{2.9$\pm$\\0.4\%} & \makecell{12.3$\pm$\\0.5\%} & \makecell{17.3$\pm$\\0.7\%} & \makecell{16.2$\pm$\\0.6\%} & \makecell{14.9$\pm$\\0.7\%} & \makecell{15.5$\pm$\\0.5\%} & \makecell{40.7$\pm$\\\ns{4.4}} & \makecell{47.2$\pm$\\\ns{2.9}} & \makecell{17.5$\pm$\\\ns{0.6}} & \makecell{14.1$\pm$\\\ns{0.4}} & \makecell{11.5$\pm$\\\ns{0.6}} & \makecell{18.3$\pm$\\\ns{1.6}} & \makecell{13.2$\pm$\\\ns{0.2}}\\ 
Mycophenolic acid & 3.20\tnote{2} & \makecell{3.3$\pm$\\0.4\%} & \makecell{2.9$\pm$\\0.3\%} & \makecell{9.8$\pm$\\0.4\%} & \makecell{12.2$\pm$\\0.5\%} & \makecell{11.8$\pm$\\0.3\%} & \makecell{12.1$\pm$\\0.6\%} & \makecell{12.5$\pm$\\0.4\%} & \makecell{47.5$\pm$\\\ns{3.1}} & \makecell{45.2$\pm$\\\ns{2.0}} & \makecell{21.0$\pm$\\\ns{0.8}} & \makecell{19.5$\pm$\\\ns{0.6}} & \makecell{15.1$\pm$\\\ns{0.6}} & \makecell{21.7$\pm$\\\ns{1.7}} & \makecell{15.7$\pm$\\\ns{0.3}}\\ 
Nafamostat & 2.00\tnote{2} & \makecell{5.8$\pm$\\0.4\%} & \makecell{5.2$\pm$\\0.4\%} & \makecell{10.4$\pm$\\0.2\%} & \makecell{12.0$\pm$\\0.3\%} & \makecell{12.0$\pm$\\0.2\%} & \makecell{12.0$\pm$\\0.4\%} & \makecell{12.3$\pm$\\0.3\%} & \makecell{26.9$\pm$\\\ns{1.7}} & \makecell{25.5$\pm$\\\ns{0.7}} & \makecell{19.8$\pm$\\\ns{1.0}} & \makecell{20.0$\pm$\\\ns{0.8}} & \makecell{15.0$\pm$\\\ns{0.5}} & \makecell{22.0$\pm$\\\ns{2.4}} & \makecell{16.1$\pm$\\\ns{0.6}}\\ 
PB28 & 5.40\tnote{2} & \makecell{4.2$\pm$\\0.4\%} & \makecell{3.6$\pm$\\0.3\%} & \makecell{9.2$\pm$\\0.3\%} & \makecell{11.0$\pm$\\0.3\%} & \makecell{10.9$\pm$\\0.2\%} & \makecell{10.6$\pm$\\0.4\%} & \makecell{11.2$\pm$\\0.3\%} & \makecell{37.7$\pm$\\\ns{1.8}} & \makecell{36.0$\pm$\\\ns{1.0}} & \makecell{22.4$\pm$\\\ns{1.1}} & \makecell{21.6$\pm$\\\ns{0.8}} & \makecell{16.4$\pm$\\\ns{0.6}} & \makecell{25.0$\pm$\\\ns{2.6}} & \makecell{17.7$\pm$\\\ns{0.6}}\\ 
PD-144418 & 3.50\tnote{2} & \makecell{3.4$\pm$\\0.3\%} & \makecell{3.0$\pm$\\0.3\%} & \makecell{8.4$\pm$\\0.3\%} & \makecell{10.2$\pm$\\0.4\%} & \makecell{10.1$\pm$\\0.2\%} & \makecell{10.3$\pm$\\0.6\%} & \makecell{10.5$\pm$\\0.3\%} & \makecell{45.8$\pm$\\\ns{2.2}} & \makecell{43.3$\pm$\\\ns{1.3}} & \makecell{24.8$\pm$\\\ns{1.0}} & \makecell{23.6$\pm$\\\ns{0.8}} & \makecell{17.8$\pm$\\\ns{0.7}} & \makecell{25.7$\pm$\\\ns{2.1}} & \makecell{19.0$\pm$\\\ns{0.5}}\\ 
PS3061 & 7.29\tnote{2} & \makecell{3.3$\pm$\\0.3\%} & \makecell{2.7$\pm$\\0.2\%} & \makecell{7.2$\pm$\\0.2\%} & \makecell{8.8$\pm$\\0.4\%} & \makecell{8.5$\pm$\\0.2\%} & \makecell{8.6$\pm$\\0.4\%} & \makecell{8.8$\pm$\\0.3\%} & \makecell{47.0$\pm$\\\ns{2.9}} & \makecell{48.1$\pm$\\\ns{1.5}} & \makecell{28.9$\pm$\\\ns{1.1}} & \makecell{27.5$\pm$\\\ns{0.8}} & \makecell{21.2$\pm$\\\ns{0.9}} & \makecell{31.3$\pm$\\\ns{2.7}} & \makecell{22.7$\pm$\\\ns{0.6}}\\ 
Pevonedistat & 1.70\tnote{2} & \makecell{3.7$\pm$\\0.4\%} & \makecell{3.2$\pm$\\0.3\%} & \makecell{9.1$\pm$\\0.3\%} & \makecell{11.3$\pm$\\0.4\%} & \makecell{11.1$\pm$\\0.3\%} & \makecell{10.6$\pm$\\0.4\%} & \makecell{11.3$\pm$\\0.3\%} & \makecell{43.1$\pm$\\\ns{2.2}} & \makecell{42.1$\pm$\\\ns{1.4}} & \makecell{23.2$\pm$\\\ns{1.0}} & \makecell{21.4$\pm$\\\ns{0.7}} & \makecell{16.5$\pm$\\\ns{0.6}} & \makecell{25.6$\pm$\\\ns{2.5}} & \makecell{17.9$\pm$\\\ns{0.5}}\\ 
Plitidepsin & 5.70\tnote{2} & \makecell{3.8$\pm$\\0.4\%} & \makecell{3.2$\pm$\\0.3\%} & \makecell{9.4$\pm$\\0.3\%} & \makecell{11.7$\pm$\\0.4\%} & \makecell{11.3$\pm$\\0.3\%} & \makecell{11.2$\pm$\\0.4\%} & \makecell{11.5$\pm$\\0.3\%} & \makecell{41.1$\pm$\\\ns{2.1}} & \makecell{40.3$\pm$\\\ns{1.3}} & \makecell{22.0$\pm$\\\ns{0.9}} & \makecell{20.3$\pm$\\\ns{0.6}} & \makecell{15.9$\pm$\\\ns{0.7}} & \makecell{23.6$\pm$\\\ns{2.3}} & \makecell{17.1$\pm$\\\ns{0.4}}\\ 
Ponatinib & 4.10\tnote{2} & \makecell{3.4$\pm$\\0.4\%} & \makecell{3.0$\pm$\\0.3\%} & \makecell{8.6$\pm$\\0.3\%} & \makecell{10.5$\pm$\\0.3\%} & \makecell{10.3$\pm$\\0.2\%} & \makecell{9.8$\pm$\\0.4\%} & \makecell{10.6$\pm$\\0.3\%} & \makecell{45.9$\pm$\\\ns{2.1}} & \makecell{43.1$\pm$\\\ns{1.4}} & \makecell{24.1$\pm$\\\ns{1.1}} & \makecell{22.9$\pm$\\\ns{0.8}} & \makecell{17.4$\pm$\\\ns{0.6}} & \makecell{27.2$\pm$\\\ns{2.7}} & \makecell{18.8$\pm$\\\ns{0.6}}\\ 
Progesterone & 3.87\tnote{1} & \makecell{3.9$\pm$\\0.4\%} & \makecell{3.3$\pm$\\0.3\%} & \makecell{10.4$\pm$\\0.3\%} & \makecell{13.0$\pm$\\0.5\%} & \makecell{12.6$\pm$\\0.3\%} & \makecell{12.0$\pm$\\0.6\%} & \makecell{12.9$\pm$\\0.4\%} & \makecell{40.6$\pm$\\\ns{2.2}} & \makecell{39.5$\pm$\\\ns{1.4}} & \makecell{19.7$\pm$\\\ns{0.8}} & \makecell{18.2$\pm$\\\ns{0.6}} & \makecell{14.1$\pm$\\\ns{0.6}} & \makecell{21.9$\pm$\\\ns{1.8}} & \makecell{15.1$\pm$\\\ns{0.4}}\\ 
RS-PPCC & 3.50\tnote{2} & \makecell{1.9$\pm$\\0.3\%} & \makecell{1.5$\pm$\\0.2\%} & \makecell{6.8$\pm$\\0.3\%} & \makecell{9.2$\pm$\\0.5\%} & \makecell{8.6$\pm$\\0.3\%} & \makecell{8.8$\pm$\\0.6\%} & \makecell{8.9$\pm$\\0.4\%} & \makecell{81.2$\pm$\\\ns{8.9}} & \makecell{87.3$\pm$\\\ns{5.2}} & \makecell{30.7$\pm$\\\ns{1.0}} & \makecell{26.2$\pm$\\\ns{0.8}} & \makecell{21.0$\pm$\\\ns{1.1}} & \makecell{30.4$\pm$\\\ns{2.1}} & \makecell{22.6$\pm$\\\ns{0.3}}\\ 
RVX-208 & 2.30\tnote{2} & \makecell{4.0$\pm$\\0.4\%} & \makecell{3.4$\pm$\\0.3\%} & \makecell{8.6$\pm$\\0.2\%} & \makecell{10.1$\pm$\\0.4\%} & \makecell{10.1$\pm$\\0.2\%} & \makecell{10.5$\pm$\\0.5\%} & \makecell{10.4$\pm$\\0.3\%} & \makecell{39.3$\pm$\\\ns{2.3}} & \makecell{38.5$\pm$\\\ns{1.2}} & \makecell{24.3$\pm$\\\ns{1.0}} & \makecell{23.8$\pm$\\\ns{0.8}} & \makecell{17.9$\pm$\\\ns{0.7}} & \makecell{25.3$\pm$\\\ns{2.3}} & \makecell{19.1$\pm$\\\ns{0.5}}\\ 
Rapamycin & 6.00\tnote{2} & \makecell{0.9$\pm$\\0.2\%} & \makecell{0.5$\pm$\\0.1\%} & \makecell{3.8$\pm$\\0.3\%} & \makecell{5.8$\pm$\\0.4\%} & \makecell{5.2$\pm$\\0.3\%} & \makecell{5.4$\pm$\\0.4\%} & \makecell{5.2$\pm$\\0.3\%} & \makecell{183.3$\pm$\\\ns{37.7}} & \makecell{246.4$\pm$\\\ns{27.5}} & \makecell{56.1$\pm$\\\ns{2.3}} & \makecell{42.4$\pm$\\\ns{1.6}} & \makecell{35.6$\pm$\\\ns{2.7}} & \makecell{50.4$\pm$\\\ns{3.0}} & \makecell{39.3$\pm$\\\ns{0.8}}\\ 
Remdesivir & 1.90\tnote{2} & \makecell{2.9$\pm$\\0.4\%} & \makecell{2.4$\pm$\\0.3\%} & \makecell{8.3$\pm$\\0.3\%} & \makecell{10.7$\pm$\\0.4\%} & \makecell{10.2$\pm$\\0.3\%} & \makecell{10.1$\pm$\\0.5\%} & \makecell{10.4$\pm$\\0.3\%} & \makecell{55.2$\pm$\\\ns{3.8}} & \makecell{56.2$\pm$\\\ns{2.4}} & \makecell{25.4$\pm$\\\ns{0.9}} & \makecell{22.6$\pm$\\\ns{0.7}} & \makecell{17.9$\pm$\\\ns{0.9}} & \makecell{26.6$\pm$\\\ns{2.2}} & \makecell{19.3$\pm$\\\ns{0.4}}\\ 
Ribavirin & -1.85\tnote{1} & \makecell{4.5$\pm$\\0.6\%} & \makecell{3.6$\pm$\\0.4\%} & \makecell{12.0$\pm$\\0.5\%} & \makecell{15.7$\pm$\\0.7\%} & \makecell{15.1$\pm$\\0.5\%} & \makecell{14.6$\pm$\\0.7\%} & \makecell{15.0$\pm$\\0.5\%} & \makecell{2.5$\pm$\\\us{0.2}} & \makecell{2.6$\pm$\\\us{0.1}} & \makecell{1.3$\pm$\\\us{0.0}} & \makecell{1.2$\pm$\\\us{0.0}} & \makecell{898.3$\pm$\\\ns{41.6}} & \makecell{1.4$\pm$\\\us{0.1}} & \makecell{1.0$\pm$\\\us{0.0}}\\ 
Ruxolitinib & 2.10\tnote{2} & \makecell{2.8$\pm$\\0.4\%} & \makecell{2.1$\pm$\\0.3\%} & \makecell{8.8$\pm$\\0.4\%} & \makecell{11.7$\pm$\\0.5\%} & \makecell{11.1$\pm$\\0.4\%} & \makecell{10.8$\pm$\\0.6\%} & \makecell{11.2$\pm$\\0.4\%} & \makecell{57.3$\pm$\\\ns{5.2}} & \makecell{61.9$\pm$\\\ns{3.3}} & \makecell{23.7$\pm$\\\ns{0.8}} & \makecell{20.5$\pm$\\\ns{0.6}} & \makecell{16.3$\pm$\\\ns{0.8}} & \makecell{24.5$\pm$\\\ns{1.8}} & \makecell{17.8$\pm$\\\ns{0.3}}\\ 
S-verapamil & 3.80\tnote{2} & \makecell{3.1$\pm$\\0.4\%} & \makecell{2.5$\pm$\\0.3\%} & \makecell{9.1$\pm$\\0.3\%} & \makecell{11.6$\pm$\\0.5\%} & \makecell{11.2$\pm$\\0.3\%} & \makecell{11.1$\pm$\\0.5\%} & \makecell{11.5$\pm$\\0.4\%} & \makecell{51.2$\pm$\\\ns{3.7}} & \makecell{52.3$\pm$\\\ns{2.4}} & \makecell{22.6$\pm$\\\ns{0.8}} & \makecell{20.4$\pm$\\\ns{0.6}} & \makecell{16.0$\pm$\\\ns{0.7}} & \makecell{23.6$\pm$\\\ns{2.1}} & \makecell{17.2$\pm$\\\ns{0.4}}\\ 
Sanglifehrin A & 7.30\tnote{2} & \makecell{1.8$\pm$\\0.2\%} & \makecell{1.5$\pm$\\0.2\%} & \makecell{5.0$\pm$\\0.2\%} & \makecell{6.6$\pm$\\0.3\%} & \makecell{6.3$\pm$\\0.2\%} & \makecell{6.4$\pm$\\0.3\%} & \makecell{6.5$\pm$\\0.2\%} & \makecell{89.2$\pm$\\\ns{6.1}} & \makecell{89.8$\pm$\\\ns{3.7}} & \makecell{41.9$\pm$\\\ns{1.5}} & \makecell{37.2$\pm$\\\ns{1.2}} & \makecell{29.2$\pm$\\\ns{1.5}} & \makecell{42.4$\pm$\\\ns{3.7}} & \makecell{31.3$\pm$\\\ns{0.6}}\\ 
Sapanisertib & 1.70\tnote{2} & \makecell{4.2$\pm$\\0.4\%} & \makecell{3.6$\pm$\\0.3\%} & \makecell{9.8$\pm$\\0.3\%} & \makecell{11.8$\pm$\\0.5\%} & \makecell{11.4$\pm$\\0.3\%} & \makecell{11.9$\pm$\\0.6\%} & \makecell{12.1$\pm$\\0.4\%} & \makecell{38.1$\pm$\\\ns{2.1}} & \makecell{37.4$\pm$\\\ns{1.2}} & \makecell{21.4$\pm$\\\ns{0.7}} & \makecell{20.5$\pm$\\\ns{0.6}} & \makecell{16.0$\pm$\\\ns{0.6}} & \makecell{22.5$\pm$\\\ns{1.8}} & \makecell{16.6$\pm$\\\ns{0.4}}\\ 
Selinexor & 3.00\tnote{2} & \makecell{2.3$\pm$\\0.4\%} & \makecell{1.8$\pm$\\0.2\%} & \makecell{7.6$\pm$\\0.4\%} & \makecell{10.3$\pm$\\0.6\%} & \makecell{9.5$\pm$\\0.4\%} & \makecell{9.8$\pm$\\0.6\%} & \makecell{9.9$\pm$\\0.4\%} & \makecell{69.9$\pm$\\\ns{7.5}} & \makecell{74.7$\pm$\\\ns{4.4}} & \makecell{27.4$\pm$\\\ns{0.9}} & \makecell{23.2$\pm$\\\ns{0.8}} & \makecell{18.9$\pm$\\\ns{1.1}} & \makecell{27.0$\pm$\\\ns{1.9}} & \makecell{20.1$\pm$\\\ns{0.3}}\\ 
Silmitasertib & 4.40\tnote{2} & \makecell{4.8$\pm$\\0.6\%} & \makecell{3.8$\pm$\\0.4\%} & \makecell{11.8$\pm$\\0.3\%} & \makecell{15.2$\pm$\\0.5\%} & \makecell{14.4$\pm$\\0.4\%} & \makecell{14.5$\pm$\\0.6\%} & \makecell{14.4$\pm$\\0.4\%} & \makecell{32.7$\pm$\\\ns{2.3}} & \makecell{34.8$\pm$\\\ns{1.5}} & \makecell{17.2$\pm$\\\ns{0.7}} & \makecell{15.3$\pm$\\\ns{0.5}} & \makecell{12.2$\pm$\\\ns{0.5}} & \makecell{17.8$\pm$\\\ns{1.6}} & \makecell{13.5$\pm$\\\ns{0.4}}\\ 
TMCB & 4.40\tnote{2} & \makecell{2.3$\pm$\\0.5\%} & \makecell{1.9$\pm$\\0.3\%} & \makecell{8.8$\pm$\\0.4\%} & \makecell{11.2$\pm$\\0.6\%} & \makecell{10.9$\pm$\\0.3\%} & \makecell{11.6$\pm$\\0.6\%} & \makecell{11.4$\pm$\\0.5\%} & \makecell{69.2$\pm$\\\ns{9.0}} & \makecell{68.5$\pm$\\\ns{5.0}} & \makecell{23.6$\pm$\\\ns{0.8}} & \makecell{21.2$\pm$\\\ns{0.8}} & \makecell{16.4$\pm$\\\ns{0.7}} & \makecell{22.7$\pm$\\\ns{1.8}} & \makecell{17.3$\pm$\\\ns{0.2}}\\ 
Ternatin 4 & 4.40\tnote{2} & \makecell{3.1$\pm$\\0.4\%} & \makecell{2.8$\pm$\\0.3\%} & \makecell{8.6$\pm$\\0.3\%} & \makecell{10.5$\pm$\\0.4\%} & \makecell{10.2$\pm$\\0.3\%} & \makecell{10.2$\pm$\\0.4\%} & \makecell{10.7$\pm$\\0.3\%} & \makecell{49.9$\pm$\\\ns{2.5}} & \makecell{47.2$\pm$\\\ns{1.7}} & \makecell{24.1$\pm$\\\ns{0.9}} & \makecell{22.8$\pm$\\\ns{0.7}} & \makecell{17.6$\pm$\\\ns{0.8}} & \makecell{26.1$\pm$\\\ns{2.4}} & \makecell{18.5$\pm$\\\ns{0.4}}\\ 
Tigecycline & 1.10\tnote{2} & \makecell{2.3$\pm$\\0.3\%} & \makecell{2.0$\pm$\\0.2\%} & \makecell{7.2$\pm$\\0.3\%} & \makecell{9.2$\pm$\\0.4\%} & \makecell{8.9$\pm$\\0.2\%} & \makecell{8.2$\pm$\\0.4\%} & \makecell{9.3$\pm$\\0.3\%} & \makecell{74.0$\pm$\\\ns{4.8}} & \makecell{71.3$\pm$\\\ns{3.2}} & \makecell{31.2$\pm$\\\ns{1.3}} & \makecell{28.5$\pm$\\\ns{0.9}} & \makecell{21.9$\pm$\\\ns{0.9}} & \makecell{35.7$\pm$\\\ns{3.3}} & \makecell{23.4$\pm$\\\ns{0.6}}\\ 
Tomivosertib & 1.30\tnote{2} & \makecell{3.3$\pm$\\0.5\%} & \makecell{2.8$\pm$\\0.3\%} & \makecell{10.1$\pm$\\0.4\%} & \makecell{12.5$\pm$\\0.6\%} & \makecell{12.0$\pm$\\0.3\%} & \makecell{12.6$\pm$\\0.6\%} & \makecell{12.7$\pm$\\0.4\%} & \makecell{50.5$\pm$\\\ns{3.6}} & \makecell{49.2$\pm$\\\ns{2.4}} & \makecell{21.4$\pm$\\\ns{0.8}} & \makecell{20.0$\pm$\\\ns{0.6}} & \makecell{15.6$\pm$\\\ns{0.6}} & \makecell{21.9$\pm$\\\ns{1.8}} & \makecell{16.2$\pm$\\\ns{0.3}}\\ 
UCPH-101 & 4.60\tnote{2} & \makecell{1.9$\pm$\\0.3\%} & \makecell{1.5$\pm$\\0.2\%} & \makecell{6.5$\pm$\\0.3\%} & \makecell{8.7$\pm$\\0.4\%} & \makecell{8.3$\pm$\\0.3\%} & \makecell{8.5$\pm$\\0.5\%} & \makecell{8.5$\pm$\\0.3\%} & \makecell{81.3$\pm$\\\ns{8.3}} & \makecell{88.5$\pm$\\\ns{5.1}} & \makecell{32.1$\pm$\\\ns{1.0}} & \makecell{27.7$\pm$\\\ns{0.9}} & \makecell{21.7$\pm$\\\ns{1.0}} & \makecell{31.5$\pm$\\\ns{2.3}} & \makecell{23.7$\pm$\\\ns{0.4}}\\ 
Valproic Acid & 2.75\tnote{1} & \makecell{7.0$\pm$\\0.7\%} & \makecell{6.2$\pm$\\0.6\%} & \makecell{16.2$\pm$\\0.5\%} & \makecell{19.6$\pm$\\0.8\%} & \makecell{19.3$\pm$\\0.4\%} & \makecell{19.9$\pm$\\0.7\%} & \makecell{19.7$\pm$\\0.6\%} & \makecell{22.2$\pm$\\\ns{1.1}} & \makecell{21.1$\pm$\\\ns{0.7}} & \makecell{12.3$\pm$\\\ns{0.5}} & \makecell{11.5$\pm$\\\ns{0.4}} & \makecell{8.9$\pm$\\\ns{0.3}} & \makecell{12.5$\pm$\\\ns{1.2}} & \makecell{9.6$\pm$\\\ns{0.2}}\\ 
Verdinexor & 4.10\tnote{2} & \makecell{2.4$\pm$\\0.4\%} & \makecell{1.8$\pm$\\0.2\%} & \makecell{8.0$\pm$\\0.4\%} & \makecell{11.0$\pm$\\0.6\%} & \makecell{10.1$\pm$\\0.4\%} & \makecell{10.2$\pm$\\0.6\%} & \makecell{10.4$\pm$\\0.4\%} & \makecell{66.6$\pm$\\\ns{7.5}} & \makecell{73.4$\pm$\\\ns{4.6}} & \makecell{25.8$\pm$\\\ns{0.8}} & \makecell{21.8$\pm$\\\ns{0.7}} & \makecell{17.9$\pm$\\\ns{1.0}} & \makecell{25.9$\pm$\\\ns{1.9}} & \makecell{19.1$\pm$\\\ns{0.3}}\\ 
WDB002 & 6.79\tnote{2} & \makecell{1.8$\pm$\\0.3\%} & \makecell{1.5$\pm$\\0.2\%} & \makecell{5.6$\pm$\\0.3\%} & \makecell{7.5$\pm$\\0.4\%} & \makecell{7.1$\pm$\\0.3\%} & \makecell{7.3$\pm$\\0.5\%} & \makecell{7.3$\pm$\\0.3\%} & \makecell{86.3$\pm$\\\ns{6.7}} & \makecell{87.7$\pm$\\\ns{4.2}} & \makecell{37.2$\pm$\\\ns{1.2}} & \makecell{32.5$\pm$\\\ns{1.0}} & \makecell{25.8$\pm$\\\ns{1.4}} & \makecell{36.8$\pm$\\\ns{2.6}} & \makecell{27.5$\pm$\\\ns{0.5}}\\ 
XL413 & 3.29\tnote{2} & \makecell{5.2$\pm$\\0.6\%} & \makecell{4.1$\pm$\\0.4\%} & \makecell{12.1$\pm$\\0.4\%} & \makecell{15.4$\pm$\\0.5\%} & \makecell{14.8$\pm$\\0.4\%} & \makecell{14.8$\pm$\\0.7\%} & \makecell{14.8$\pm$\\0.4\%} & \makecell{30.2$\pm$\\\ns{2.0}} & \makecell{31.6$\pm$\\\ns{1.1}} & \makecell{16.8$\pm$\\\ns{0.7}} & \makecell{15.0$\pm$\\\ns{0.5}} & \makecell{11.8$\pm$\\\ns{0.5}} & \makecell{17.3$\pm$\\\ns{1.5}} & \makecell{13.1$\pm$\\\ns{0.3}}\\ 
ZINC1775962367 & 4.44\tnote{2} & \makecell{1.9$\pm$\\0.4\%} & \makecell{1.4$\pm$\\0.2\%} & \makecell{6.5$\pm$\\0.3\%} & \makecell{9.2$\pm$\\0.4\%} & \makecell{8.5$\pm$\\0.4\%} & \makecell{8.5$\pm$\\0.5\%} & \makecell{8.4$\pm$\\0.3\%} & \makecell{82.2$\pm$\\\ns{9.4}} & \makecell{94.3$\pm$\\\ns{5.8}} & \makecell{32.2$\pm$\\\ns{1.0}} & \makecell{26.2$\pm$\\\ns{0.8}} & \makecell{21.2$\pm$\\\ns{1.3}} & \makecell{31.6$\pm$\\\ns{2.4}} & \makecell{23.9$\pm$\\\ns{0.4}}\\ 
ZINC4326719 & 2.60\tnote{2} & \makecell{3.3$\pm$\\0.5\%} & \makecell{2.4$\pm$\\0.3\%} & \makecell{8.6$\pm$\\0.3\%} & \makecell{11.5$\pm$\\0.4\%} & \makecell{10.8$\pm$\\0.4\%} & \makecell{10.9$\pm$\\0.6\%} & \makecell{10.8$\pm$\\0.4\%} & \makecell{48.0$\pm$\\\ns{4.0}} & \makecell{55.4$\pm$\\\ns{2.4}} & \makecell{24.1$\pm$\\\ns{0.8}} & \makecell{20.7$\pm$\\\ns{0.6}} & \makecell{16.7$\pm$\\\ns{0.8}} & \makecell{24.3$\pm$\\\ns{1.9}} & \makecell{18.3$\pm$\\\ns{0.4}}\\ 
ZINC4511851 & 3.41\tnote{2} & \makecell{4.6$\pm$\\0.5\%} & \makecell{3.8$\pm$\\0.3\%} & \makecell{10.8$\pm$\\0.3\%} & \makecell{13.3$\pm$\\0.5\%} & \makecell{12.9$\pm$\\0.3\%} & \makecell{13.0$\pm$\\0.6\%} & \makecell{13.2$\pm$\\0.4\%} & \makecell{33.9$\pm$\\\ns{1.9}} & \makecell{34.1$\pm$\\\ns{1.1}} & \makecell{18.9$\pm$\\\ns{0.7}} & \makecell{17.7$\pm$\\\ns{0.5}} & \makecell{13.7$\pm$\\\ns{0.6}} & \makecell{20.0$\pm$\\\ns{1.7}} & \makecell{14.8$\pm$\\\ns{0.4}}\\ 
ZINC95559591 & 5.33\tnote{2} & \makecell{4.5$\pm$\\0.4\%} & \makecell{4.1$\pm$\\0.4\%} & \makecell{11.0$\pm$\\0.3\%} & \makecell{13.0$\pm$\\0.4\%} & \makecell{12.9$\pm$\\0.2\%} & \makecell{12.1$\pm$\\0.4\%} & \makecell{13.5$\pm$\\0.3\%} & \makecell{35.1$\pm$\\\ns{1.3}} & \makecell{32.1$\pm$\\\ns{1.0}} & \makecell{18.5$\pm$\\\ns{0.9}} & \makecell{18.1$\pm$\\\ns{0.6}} & \makecell{13.7$\pm$\\\ns{0.5}} & \makecell{21.6$\pm$\\\ns{2.3}} & \makecell{14.4$\pm$\\\ns{0.5}}\\ 
Zotatifin & 2.40\tnote{2} & \makecell{2.7$\pm$\\0.4\%} & \makecell{2.1$\pm$\\0.3\%} & \makecell{8.4$\pm$\\0.3\%} & \makecell{10.9$\pm$\\0.4\%} & \makecell{10.3$\pm$\\0.4\%} & \makecell{10.0$\pm$\\0.5\%} & \makecell{10.6$\pm$\\0.4\%} & \makecell{58.4$\pm$\\\ns{4.9}} & \makecell{61.9$\pm$\\\ns{3.2}} & \makecell{24.8$\pm$\\\ns{0.9}} & \makecell{22.0$\pm$\\\ns{0.7}} & \makecell{17.4$\pm$\\\ns{0.9}} & \makecell{26.6$\pm$\\\ns{2.2}} & \makecell{18.8$\pm$\\\ns{0.4}}\\ 
dBET6 & 5.00\tnote{2} & \makecell{1.9$\pm$\\0.3\%} & \makecell{1.5$\pm$\\0.2\%} & \makecell{5.4$\pm$\\0.2\%} & \makecell{7.1$\pm$\\0.3\%} & \makecell{6.8$\pm$\\0.2\%} & \makecell{7.0$\pm$\\0.4\%} & \makecell{6.9$\pm$\\0.3\%} & \makecell{82.9$\pm$\\\ns{6.2}} & \makecell{86.2$\pm$\\\ns{3.9}} & \makecell{38.8$\pm$\\\ns{1.3}} & \makecell{34.5$\pm$\\\ns{1.1}} & \makecell{27.0$\pm$\\\ns{1.3}} & \makecell{38.8$\pm$\\\ns{3.2}} & \makecell{29.3$\pm$\\\ns{0.6}}\\ 
  \bottomrule
  \end{longtable}
  \begin{tablenotes}
  \small

  \item[1] These partition coefficients were taken from PubChem.
  \item[2] These partition coefficients were taken from XLogP3 model~\textsuperscript{38}.
  \item[3] parameters for plasma membrane.
  \item[4] parameters for lysosome membrane.
  \item[5] parameters for Golgi membrane.
  \item[6] parameters for mitochondrial membrane.
  \item[7] parameters for endoplasmic reticulum membrane.
  \item[8] parameters for coronavirus membrane with spike protein.
  \item[9] parameters for coronavirus membrane without spike protein.
  
  \end{tablenotes}
\end{ThreePartTable}

\end{landscape}

\clearpage